\documentclass[aps,prd,floatfix,nofootinbib,superscriptaddress,twocolumn,showkeys,10pt]{revtex4-2}
\usepackage{placeins}

\usepackage{amsmath}

\usepackage{amssymb}
\allowdisplaybreaks

\usepackage{graphicx}
\usepackage{tikzit}

\tikzstyle{new style 0}=[fill=black, draw=none, shape=circle]

\tikzstyle{dashed edge 1}=[-, dashed]
\tikzstyle{dashed edge 2}=[->, dashed]

\usepackage{float}

\usepackage{makeidx}
\usepackage{color}
\usepackage{amsfonts}
\usepackage[usenames,dvipsnames]{pstricks}

\usepackage{epsfig}
\usepackage{pst-grad} 
\usepackage{mathtools}
\usepackage[colorlinks,hyperindex]{hyperref}
\usepackage{array}

\hypersetup
{	colorlinks,%
	citecolor=black,%
	linkcolor=black,%
	urlcolor=black,%
}

\usepackage{allpurpose}

\usepackage{flushend}

\newcommand\nn{{\nonumber}}
\def\slashi#1{\rlap{\sl/}#1}
\def\rme{\mathrm{e}}

\usepackage{ulem}
\usepackage{color}

\begin{document}

\title{From Matter Density to Deflection Angle and Gravitational Lensing Using a Perturbative Method}

\author{Peiran Liu}
\affiliation{School of Physics and Technology, Wuhan University, Wuhan, 430072, China}

\author{Xiaotian Zhang}
\affiliation{School of Physics and Technology, Wuhan University, Wuhan, 430072, China}

\author{Junji Jia}
\email[Corresponding author:~]{junjijia@whu.edu.cn}
\affiliation{Department of Astronomy \& MOE Key Laboratory of Artificial Micro- and Nano-structures, School of Physics and Technology, Wuhan University, Wuhan, 430072, China}

\date{\today}

\begin{abstract}
In this work, we develop a perturbative method to compute the deflection angle of null or timelike signals in spacetimes filled with a static and spherically symmetric (SSS) perfect fluid with fairly arbitrary density distributions. After solving the Tolman-Oppenheimer-Volkoff equations, the metric functions of the spacetime are obtained either as asymptotic series or as expansions around a finite boundary. The deflection angles of null or timelike signals in the weak-field limit in such spacetimes can then be expressed as series expansions in terms of the impact parameter, with coefficients determined by the metric expansions and, in turn, the density distribution function. Gravitational lensing equations are also solved perturbatively to derive the apparent angles of the lensed images. Comparing our perturbative formulas with numerical results demonstrates the validity and efficiency of our method and results.
This procedure establishes a direct connection between the mass density, the deflection angle, and the apparent angles of gravitationally lensed images. We apply these methods and results to the generalized Navarro-Frenk-White model and some other density profiles to analyze the influence of the density parameters.  
\end{abstract}


\maketitle
\section{Introduction\label{secmmmintro}}

The deflection of light rays was one of the most significant pieces of evidence supporting the validity of General Relativity (GR) \cite{einstein1936lens,dyson1920ix}.
The value of the deflection angle around the Sun was initially derived under the assumption that the spacetime exhibits Schwarzschild-like properties. In contemporary astronomy, the bending of light rays by more massive systems, such as galaxies or galaxy clusters, is known as gravitational lensing (GL). It has developed into a crucial observational technique.  
GL has been used in a wide range of applications, including measuring the mass profile of galaxies (or galaxy clusters) \cite{bartelmann2001weak,planck2016planck,refregier2003weak,lewis2006weak,kochanek1991implications}, determining the Hubble constant \cite{bartelmann2001weak,refsdal1964possibility,keeton1997determining}, deriving cosmological parameters \cite{falco1998limits,bartelmann1999power}, and detecting dark matter \cite{bartelmann2001weak,bartelmann1999power,maoz1993early,kneib1996hubble,hoekstra1998weak,metcalf2001compound}. More recently, it has been utilized to test the speed and frequency of gravitational waves \cite{stefanov2010connection}. Additionally, GL serves as a valuable tool for probing gravitational theories beyond GR, including various modified gravity theories \cite{keeton2005formalism,joyce2016dark}. 

Although the most fundamental principles of GL are often introduced in textbooks using simple static and spherically symmetric (SSS) spacetimes, typically Schwarzschild or, at most, Kerr spacetimes \cite{perlick2004gravitational,bozza2005analytic}, realistic GL phenomena occur within continuous mass distributions formed by galaxies or galaxy clusters. Previously, to investigate the qualitative properties of certain kinds of deflection and lensing, such as the deflection of timelike signals, the influence of finite distances between the source and lens, the role of electromagnetic interactions \cite{pugliese2011motion,das2017motion,pugliese2017general,abdujabbarov2010test,al2013critical,Tursunov:2016dss,Xu:2021rld,Crisnejo:2019xtp}, and the theoretical methodologies for calculating the deflection angle (such as the Gauss-Bonnet theorem-based methods and the perturbative method), numerous studies on signal deflection and GL have been conducted in idealized background spacetimes. This situation motivates us in this work to study the feasibility of analyzing some of the aforementioned effects and methodologies in signal deflections and GL directly from relativistic mass distributions.

Many complex and well-established models for galaxy or dark matter mass distributions have been proposed, such as the Navarro-Frenk-White (NFW) model introduced by Navarro, Frenk, and White in 1997 \cite{Navarro:1996gj}, and the Hernquist model, proposed by Lars Hernquist in 1990 \cite{Hernquist:1990be}. In this work, we demonstrate that the perturbative method, previously developed for simple SSS background spacetimes, can be extended to more complex mass density profiles, such as the generalized NFW (gNFW) profile, the Hernquist profile, and the power-law profile. Additionally, we show that the finite distance effect and the deflection and GL of timelike signals can also be derived simultaneously within these systems.

Since our focus is on SSS spacetimes filled with a perfect fluid, we begin by employing the Tolman-Oppenheimer-Volkoff (TOV) equations to derive the spacetime metric corresponding to a given mass profile. The subsequent step, which involves the perturbative process from the metric functions to the bending angle and GL apparent angles, has long been demonstrated to work effectively. Consequently, the most critical step is to demonstrate that the TOV equation can be correctly solved using the perturbative method. This turns out to be feasible, either asymptotically if the mass profile decreases rapidly enough with increasing radius coordinate, or via a series expansion method around a finite boundary, depending on the specific characteristics of the chosen mass models. This paper aims to solve the TOV equations perturbatively to derive series expansions for the metric functions, which are subsequently used to determine the defection angle and to solve the GL equation. The proposed approach is formulated to accommodate general density profiles for which analytical metric solutions are difficult to obtain.

The structure of this paper is as follows. In Sec. \ref{secmmmmetricandsol}, we demonstrate how to solve the general TOV equations perturbatively to obtain the metric functions for a general mass distribution, using either an asymptotic expansion or a series expansion around a finite radius. Sec. \ref{secmmmdefang} focuses on the perturbative method for determining deflection angles in terms of the expansion coefficients of the density functions. The results obtained in this section are then applied to specific mass models, including gNFW and its subclass models, the Hernquist model, the pseudo-isothermal sphere (PIS) model, and the power-law model, in Sec. \ref{secmmmappgl}. This section derives the deflection angles in each model and investigates the effect of various parameters of the matter density profiles. These deflection angles are then used in Sec. \ref{secmmmgl} to solve the GL equation for the corresponding models, yielding the apparent angles of the images. Finally, Sec. \ref{secmmmdisc} concludes the paper with a brief summary and discussion. 
Throughout the paper, we adopt natural units with $G=c=1$. 

\section{TOV equations and their solutions\label{secmmmmetricandsol}}

We start from the general line element of the SSS spacetime
\be\label{eqmmmmetric}
\mathrm{d}s^2 = -A(r)\mathrm{d}t^2+B(r)\mathrm{d}r^2+C(r)(\mathrm{d}\theta^2+\sin{\theta^2}\mathrm{d}\phi^2),
\ee
where $C(r)=r^2$ and $(t,~r,~\theta,~\phi)$ are the coordinates, while $A(r),~B(r)$ are functions that depend on $r$ only. When we study the solution of a spacetime filled with a perfect fluid characterized by a density distribution $\rho(r)$ and pressure $P(r)$, the metric functions $A(r)$ and $B(r)$ are often expressed in terms of the gravitational potential $\Phi(r)$ and the effective mass $m(r)$ enclosed within a radius $r$, as described by the following relation
\begin{subequations}
\label{eqmmmabinphim}
    \begin{align}
        &A(r) = e^{2\Phi(r)},\\
        &B(r) = \lsb1-\frac{2m(r)}{r}\rsb^{-1}.
    \end{align}
\end{subequations}
The Einstein equation for such a perfect fluid is then reduced to the well-known TOV equations \cite{Wald:1984rg}
\begin{subequations}\label{eqmmmtov}
    \begin{align}
        & \frac{\mathrm{d}P}{\mathrm{d}r} = -\left(\rho+P\right)\frac{m+4\pi r^3 P}{r\left(r-2m\right)},\label{tova} \\
        & \frac{\mathrm{d}m}{\mathrm{d}r} = 4\pi r^2 \rho,\label{tovb} \\
        & \frac{\mathrm{d}\Phi}{\mathrm{d}r} = \frac{m+4\pi r^3 P}{r\left(r-2m\right)}. \label{tovc}
    \end{align}
\end{subequations}

One of the main purposes of this work is to derive the solution to the deflection angle 
of null or timelike signals, once a density distribution and the kinetic properties of the signals are specified. To achieve this, it is first necessary to derive the explicit forms of the metric functions $A(r),~B(r)$, or equivalently $\Phi(r)$ and $m(r)$, from the given density distribution $\rho(r)$.

While deriving for $m(r)$ from $\rho(r)$ appears straightforward based on Eq. \eqref{tovb}, the explicit form of $m(r)$ depends on the integrability of $r^2\rho(r)$. In contrast, determining the pressure $P(r)$ and, consequently, the potential $\Phi(r)$, is more challenging due to the nonlinear nature of Eq. \eqref{tova}. To address these challenges, we use two series-based methods to solve for $m(r)$, $\Phi(r)$, and consequently the metric functions $A(r)$ and $B(r)$. The first method generates an asymptotic series valid for large radii $r\to\infty$, while the second method expands the solutions around a fixed boundary radius $r=R$. In this section, we derive the formal solutions for these quantities. The validity of these solutions will be demonstrated in Sec. \ref{secmmmappgl} using specific density profiles.

\subsection{Method 1: Asymptotic solution\label{subsecmmmnont}}

To derive the asymptotic solutions, we assume that the density $\rho(r)$ allows the following asymptotic expansion
\be\label{eqmmmdensityinfty1}
\rho(r)=\sum_{n=1}^{\infty}\frac{\rho_n}{r^{n+\delta}},
\ee 
where coefficients $\rho_n$ are determined by the specified mass model, as will be discussed in Sec. \ref{secmmmappgl}. Here, the parameter $\delta\in [0,1)$ in the power of $r$ is introduced to account for potential non-integer power series that may appear in certain density models (e.g., see the general three-parameter model Eq. \eqref{eqmmmmmbf}). The solution process for the corresponding TOV equations, metrics, and deflection angles will vary depending on whether $\delta$ is zero. Accordingly, we will first analyze the case $\delta=0$ and subsequently address the scenario where $\delta\in(0,1)$.

\subsubsection{Density with integer power \label{sssecmmmdensitywithip}}

Substituting Eq. \eqref{eqmmmdensityinfty1} with $\delta=0$ into Eq. \eqref{tovb}, the mass function can be obtained as
\be\label{eqmmmmassinfty1}
m(r)=M+4\pi\lsb\rho_3\ln r-\sum_{n=1,n\neq3}^{\infty}\frac{\rho_n}{\left(n-3\right)r^{n-3}}\rsb
\ee
where $M$ is an integration constant. 
Since our focus is on asymptotically flat spacetimes, which requires that $\displaystyle \lim_{r\to\infty}m(r)/r=0$, it follows that the first two coefficients $\rho_1, \rho_2$ in Eq. \eqref{eqmmmdensityinfty1} must vanish. Consequently, for $\rho(r)$ and $m(r)$, we have
\begin{align}
&\rho(r)=\sum_{n=3}^{\infty}\frac{\rho_n}{r^n},\label{eqmmmdensityinfty3}\\
&
m(r)=M+4\pi\lsb\rho_3\ln{r}-\sum_{n=4}^{\infty}\frac{\rho_n}{\left(n-3\right)r^{n-3}}\rsb. \label{eqmmmmrinfinrho}
\end{align}
Here, $M$ can be interpreted as the effective total mass as $r\to\infty$ if $\rho_3=0$. However, if $\rho_3\neq 0$, $M$ will be equal to $m_0+4\pi\rho_3\ln l$ for some length scale $l$, ensuring the cancellation of the anomalous dimensional contribution arising from the $4\pi\rho_3\ln r$ term in Eq. \eqref{eqmmmmrinfinrho}. In principle, both $m_0$ and $l$ may depend on the density coefficients $\rho_n$ and the lower limit of the integral used to compute $m(r)$. 

In order to solve Eq. \eqref{tova} for the pressure $P(r)$, we found that the method of undetermined coefficients can be applied by assuming an ansatz in the form of a double series expansion for $P(r)$ as follows
\be\label{eqmmmpressureinfty1}
P(r)=\sum_{n=4}^{\infty}\sum_{m=0}^{n-3}P_{n,m}\frac{\left(\ln{r}\right)^m}{r^n}.
\ee
The $\ln r$ terms arise primarily from the presence of the $\ln r$ in $m(r)$ in Eq. \eqref{eqmmmmrinfinrho}. If $\rho_3=0$, all terms of the form $P_{n,m>0}$ vanish. It is worth noting that the series representation in Eq. \eqref{eqmmmpressureinfty1} inherently satisfies the boundary condition $P(r\to\infty)=0$.
Substituting Eqs. \eqref{eqmmmdensityinfty3}-\eqref{eqmmmpressureinfty1} into Eq. \eqref{tova}, the coefficients $P_{n,m}$ can be determined by comparing the coefficients of each power of $\left(\ln{r}\right)^m/r^n$ on both sides of the equation. These coefficients can be expressed in terms of $\rho_n$ and the constant $M$. The first few orders are given by 
\begin{subequations}\label{eqmmmpinffirst}
    \begin{align}
P_{4,0}=&\frac{1}{4}\rho_3\left(M+\pi\rho_3\right),~P_{4,1}=\pi\rho_3^2,\\
P_{5,0}=& \rho _3 \left(\frac{9}{20} M^2+\frac{97}{100} \pi  M \rho _3+\frac{122}{125} \pi ^2 \rho _3^2-\frac{16}{25} \pi  \rho _4\right)\nonumber\\
&+\frac{M\rho_4}{5}, \\
P_{5,1}=&\frac{1}{25} \pi  \rho _3 \left(90 M \rho _3+97 \pi  \rho _3^2+20 \rho _4\right),\\
P_{5,2}=&\frac{36}{5}\pi^2\rho_3^3.
\end{align}
\end{subequations}
Higher-order terms can also be easily obtained.

For the potential $\Phi(r)$, we can simply substitute Eqs. \eqref{eqmmmmrinfinrho} and \eqref{eqmmmpressureinfty1} into Eq. \eqref{tovc} to derive its series expansion. It takes the form given in
\begin{align}
    \Phi(r)=\sum_{n=1}^{\infty}\sum_{m=0}^{n}\Phi_{n,m}\frac{\left(\ln{r}\right)^m}{r^n},
\end{align}
where $\Phi_{n,m}$ are the coefficients. Their first few orders are provided by
\begin{subequations}\label{eqmmmmphiinf}
    \begin{align}
    \Phi_{1,0}&=-M-4\pi\rho_3,~~\Phi_{1,1}=-4\pi\rho_3,\\
    \Phi_{2,0}&=\frac{1}{2}\left(-2M^2-9\pi M\rho_3-19\pi^2\rho_3^2+4\pi\rho_4\right),\\
    \Phi_{2,1}&=\frac{1}{2}\left(-16\pi M\rho_3-36\pi^2\rho_3^2\right),\\
    \Phi_{2,2}&=-16\pi^2\rho_3^2.
    \end{align}
\end{subequations}
By substituting the solution Eq. \eqref{eqmmmmphiinf}, along with the solution from Eq. \eqref{eqmmmmrinfinrho}, into Eq. \eqref{eqmmmabinphim}, we derive the asymptotic solutions of the metric functions $A(r)$ and $B(r)$ as 
\begin{subequations}\label{eqmmmabinfser}
    \begin{align}
&A(r)=1+\sum_{n=1}^{\infty}\sum_{m=0}^{n-1+\delta_{1n}}a_{n,m}\frac{\left(\ln{r}\right)^m}{r^n},\\
&B(r)=1+\sum_{n=1}^{\infty}\sum_{m=0}^n b_{n,m}\frac{\left(\ln{r}\right)^m}{r^n},
    \end{align}
\end{subequations}
where $a_{n,m}$ and $b_{n,m}$ are the coefficients, and $\delta_{1n}$ is the Kronecker delta. Their first several orders are given by
\begin{subequations}
\label{eqmmmabinffirst}
    \begin{align}
        a_{1,0}=&-2M-8\pi\rho_3,\quad &&a_{1,1}=-8\pi\rho_3,\\
        a_{2,0}=&\pi  \left(7 M \rho _3+13 \pi  \rho _3^2+4 \rho _4\right),~&&a_{2,1}=28\pi^2\rho_3^2,\\
        b_{1,0}=&2M,\quad &&b_{1,1}=8\pi\rho_3,\\
        b_{2,0}=&4M^2-8\pi\rho_4,\quad &&b_{2,1}=32\pi M\rho_3,\nonumber\\ b_{2,2}=&64\pi^2\rho_3^2.
    \end{align}
\end{subequations}
Again, higher-order terms can be derived straightforwardly. It is worth noting that all the coefficients $P_{n,m}$ in Eq. \eqref{eqmmmpinffirst}, as well as $a_{n,m},~b_{n,m}$ in Eq. \eqref{eqmmmabinffirst} for $m>1$, are proportional to at least first order of $\rho_3$. This indicates that these coefficients vanish when $\rho_3=0$, as expected. The parameter $\rho_3=0$ eliminates the logarithmic term in $m(r)$, which, in turn, removes the corresponding contributions in $P(r),~A(r)$ and $B(r)$. 

\subsubsection{Density with non-integer power}\label{sssecmmmdensitywithfp}

When the density profile in Eq. \eqref{eqmmmdensityinfty1} takes the form of a non-integer power series, as in the case of the general three-parameter model in Eq. \eqref{eqmmmmmbf} for certain choices of parameters $(\alpha,\beta,\gamma)$, the solution for $m(r),~P(r)$ and $A(r),~B(r)$ will differ slightly.

For the spacetime to be asymptotically flat in this case, $\rho_1$ in Eq. \eqref{eqmmmdensityinfty1} must still be zero, while $\rho_2$ may now be nonzero. Furthermore, when $\delta$ is an irrational number, solving the equations using the perturbative method becomes highly complicated. Therefore, in this work, we assume that $\delta=s/t $ is a rational number in its reduced form, i.e., $s,~t\in \mathbb{Z}_>,~s<t,~\mathrm{GCD}(s,t)=1$. Consequently, we will begin with the density
\begin{align}
\label{eqmmmdensityinftyd}
\rho(r)=\sum_{n=2}^{\infty}\frac{\rho_n}{r^{n+\delta}},~~\delta=\frac{s}{t}\in\mathbb{Q},~s,t\in\mathbb{Z}_>.
\end{align}
By directly substituting this expression with a nonzero $\delta$ into Eq. \eqref{tovb}, the resulting mass profile takes the form of 
\begin{align}
\label{eqmmmnonintegerm}
m(r)=M-\sum_{n=-1}^{\infty}\frac{4\pi\rho_{n+3}}{\left(n+\delta\right)r^{n+\delta}},
\end{align}
where $M$ is an integration constant. By substituting Eqs. \eqref{eqmmmdensityinftyd} and \eqref{eqmmmnonintegerm}, along with the ansatz 
\begin{align}
\label{eqmmmnonintegerP2}
P(r)=\sum_{n=2}^\infty \sum_{m=1}^{t}P_{n,m}\frac{1}{r^{n+\frac{m}{t}}}
\end{align}
into Eq. \eqref{tova}, we can solve for the coefficients $P_{n,m}$ in terms of $\rho_n,~\delta$, and the constant $M$. The forms of the first few orders are as follows
\begin{subequations}\label{eqmmmnonintegerP}
    \begin{align}
    P_{2,1}=&0,\\
    P_{2,2}=&\frac{2\pi \rho_2^2}{1-\delta^2},\\
    P_{2,k}=&-\frac{4 \pi  \left[\left(2 k-3\right)\delta  +6\right]}{\left(\delta -1\right) \left(k\delta  +2\right)}\rho_2 P_{2,k-1}\nonumber\\
    &+4\pi\sum_{m=1}^{k-1}\frac{P_{2,m}P_{2,k-m}}{2+k\delta},\qquad k=3,~\cdots,~t.
    \end{align}
\end{subequations}
Higher-order terms can also be obtained iteratively without difficulty. However, they are not shown here due to their excessive length. 

Furthermore, we can show that the metric functions $A(r)$ and $B(r)$ have forms similar to that of $P(r)$
\begin{subequations}
\begin{align} &A(r)=\sum_{n=0}^{\infty}\sum_{m=0}^{t-1}a_{n,m}\frac{1}{r^{n+\frac{m}{t}}},\label{eqmmmnonintegerA}\\
&B(r)=\sum_{n=0}^{\infty}\sum_{m=0}^{t-1}b_{n,m}\frac{1}{r^{n+\frac{m}{t}}},\label{eqmmmnonintegerB}
\end{align}
\end{subequations}
with the first few coefficients given by
\begin{subequations}\label{eqmmmnonintegerab}
    \begin{align}
        a_{0,0}=&1,\\
        a_{0,k}=&\frac{8 \pi  \rho _2 \left(\delta -k\delta+1\right) a_{0,k-1}}{k\left(\delta -1\right) \delta }-\frac{8 \pi  }{k\delta }\sum _{i=0}^{k-1} a_{0,i} P_{2,k-i},\nonumber\\
        &~~~~~~ k=1,~\cdots,~t-1,\\
        b_{0,k}=&\lb\frac{8 \pi  \rho _2}{1-\delta}\rb^k,\quad k=0,~1,~\cdots,~t-1.
    \end{align}
\end{subequations}
Again, higher-order terms are not shown here. 

\subsection{Method 2: finite boundary solution \label{subsecmmmmetricR}}

The expansion and results presented in the previous subsection are particularly useful for the matter distribution that extends to spatial infinity. However, in many astrophysical applications, it is assumed that matter exists only within a finite range, i.e., $r<R$, beyond which there is only a vacuum. For such matter distributions, the metric within the matter region is still governed by the TOV equations, while outside it must match the vacuum solution, specifically the Schwarzschild solution. Therefore, in this subsection, we establish a perturbative procedure applicable to this scenario. 

For a finite spherical distribution, if it is nonsingular at the boundary $r=R$, we can always assume that it can be expanded around $R$ as shown in
\be\label{eqmmmdensityR}
\rho(r)=\sum_{n=0}^{\infty}\rho_n\left(r-R\right)^n~~\mathrm{for}~~ r<R,
\ee
and zero for $r>R$.
Using Eq. \eqref{tovb}, we can integrate to find the mass function $m(r)$ and express it as a series expansion around $R$
\begin{align} \label{eqmmmmassR}
m(r) =& \int_0^r4\pi r'^2\rho(r')\mathrm{d}r'+C_m\nonumber\\
=&\sum_{n=0}^{\infty}m_n\left(r-R\right)^n,
\end{align}
with coefficients
\begin{subequations}\label{eqmmmmcoeffR}
    \begin{align}
        &m_0=4\pi\sum_{n=0}^{\infty}\rho_n\frac{2R^3\left(-R\right)^n}{\left(1+n\right)\left(2+n\right)\left(3+n\right)}+C_m\equiv M,\\
        &m_1=4\pi R^2\rho_0,\\
        &m_2=2\pi R\left(2\rho_0+R\rho_1\right),\\
        &m_n=\frac{4\pi}{n}\left(\rho_{n-3}+2R\rho_{n-2}+R^2\rho_{n-1}\right), ~~n\geq3.
    \end{align}
\end{subequations}
The coefficient $m_0$ in Eq. \eqref{eqmmmmcoeffR}, also denoted as $M$, represents the total mass within $r=R$. It is obtained by integrating all terms and then setting $r=R$. If there is no singularity at $r=0$, (i.e., $m(0)=0$), then the additional integration constant $C_m$ should be set to zero.

For the solution of the pressure $P(r)$, similar to the approach in Sec. \ref{subsecmmmnont}, we can apply the method of undetermined coefficients by assuming that $P(r)$ can be expressed as a series given by 
\be\label{eqmmmpressureR}
P(r)=\sum_{n=0}^{\infty}P_n\left(r-R\right)^n~~\mathrm{for}~~ r<R,
\ee
and $0$ for $r>R$. Here, we have not imposed the natural condition $P(R)=0$, which would require $P_0=0$, because certain astrophysical models, such as the singular isothermal sphere (SIS) density functions, do not adhere to this condition. 
By substituting Eqs. \eqref{eqmmmdensityR}, \eqref{eqmmmmassR} and \eqref{eqmmmpressureR} into Eq. \eqref{tova}, and comparing the coefficients of the same order, we find that each $P_n$ can be expressed as an $n$-degree polynomial in the coefficients $\{\rho_0,~\cdots,~\rho_{n-1}\}$, i.e.,
\be\label{eqmmmcoepressureR}
P_n = P_n\left(\rho_0,~\cdots, ~\rho_{n-1}\right),\qquad n=1,~2,~\cdots.
\ee
Here, $P_n$ on the right-hand side represents not only the coefficient of pressure but also the $n$-degree polynomial. Below, we present the results for $P_1$ and $P_2$ in
\begin{subequations}\label{eqmmmPcoeffR}
\begin{align}
    P_1=&\frac{\left(\rho_0+P_0\right)\left(M+4\pi R^3 P_0\right)}{R\left(2M-R\right)},\\
        P_2=&\frac{1}{2R^2\left(R-2M\right)^2}\left\{-M^2\left(P_0-2\rho_1R+\rho_0\right)\right.\nonumber\\&\left.+M \left[28\pi R^3 P_0^2+R\left(8\pi \rho_1 R^3+32\pi \rho_0 R^2+2\right)P_0\right.\right.\nonumber\\&\left.\left.+R\left(4\pi \rho_0^2R^2-\rho_1 R+2\rho_0\right)\right] +32\pi^2 R^6P_0^3\right.\nonumber\\&\left.+4\pi R^4\left(4\pi \rho_0R^2-1\right)P_0^2\right.\nonumber\\&\left.-4\pi R^4\left(4\pi \rho_0^2R^2+\rho_1 R+2\rho_0\right)P_0-4\pi \rho_0^2R^4\right\},
\end{align}
\end{subequations}
while $P_0$ can be considered an integration constant that must be determined by applying the appropriate boundary condition. 

We can then use Eq. \eqref{tovc}, along with the series solutions for $m(r)$ and $P(r)$, to solve for the potential $\Phi(r)$ as a Taylor series around $R$. 
\begin{align}\label{eqmmmphirsolR}
    \Phi(r)=\sum_{n=0}^{\infty}\Phi_{n}\left(r-R\right)^n,~~r<R,
\end{align}
with the first few coefficients given by
\begin{subequations}
\begin{align}\label{eqmmmphicoeffR}
    \Phi_0 =& \Phi_0,\\
    \Phi_1 =& \frac{M+4\pi R^3P_0}{\left(R-2M\right)R},\\
    \Phi_2 =& \frac{1}{\left(R-2M\right)^2R^2}\left[-8\pi^2R^6P_0^2\right.\nonumber\\&\left.+2\pi R^3P_0\left(4\pi R^3\rho_0+R-5M\right)\right.\nonumber\\&\left.+\left(2\pi R^3\rho_0-M\right)\left(R-M\right)\right].
\end{align}
\end{subequations}
Here, $\Phi_0$ is determined the matching condition between the perturbative interior metric and the exterior Schwarzchild metric at the boundary $r=R$. Subsequently, by substituting this into Eq. \eqref{eqmmmabinphim}, we obtain the solution for $A(r)$ and $B(r)$ in this case as
\begin{subequations}
\label{eqmmmarbrsol2}
    \begin{align}
        A(r)=& \sum_{n=0}^{\infty}a_n\left(r-R\right)^n,~~r<R,\\
        B(r)=& \sum_{n=0}^{\infty}b_n\left(r-R\right)^n,~~r<R,
    \end{align}
\end{subequations}
with the first several coefficients given by
\begin{subequations}
\label{eqmmmabcoeffR}
    \begin{align}
        a_0 =& 1-\frac{2M}{R},\\
        a_1 =& \frac{2M}{R^2}+8\pi R P_0,\\
        a_2 =& -\frac{1}{(2M-R)R^3}\left\{4M^2\right.\nonumber\\&\left.-M\lsb4\pi R^3 P_0+2R\left(2\pi \rho_0 R^2+1\right)\rsb\right.\nonumber\\&\left.+16\pi^2R^6P_0^2+4\pi R^4\left(4\pi \rho_0R^2+1\right)P_0+4\pi \rho_0 R^4\right\},\\
        b_0 =& \lb1-\frac{2M}{R}\rb^{-1},\\
        b_1 =& \frac{-2M+8\pi\rho_0 R^3}{\left(R-2M\right)^2},\\
        b_2 =& \frac{1}{\left(2M-R\right)^3}\left[M\left(8\pi \rho_1 R^3+32\pi \rho_0 R^2-2\right)\right.\nonumber\\&\left.-R^4\left(64\pi^2\rho_0^2 R+4\pi \rho_1\right)\right].
    \end{align}
\end{subequations}
Again, higher-order terms can be computed straightforwardly. 

\section{The deflection angle 
\label{secmmmdefang}}

Once the metric functions are determined, we can calculate the deflection angle 
of a signal traveling from the source to the detector by integrating the geodesic equations. 

The geodesic equations in the spacetime described by the metric Eq.
\eqref{eqmmmmetric} are given by
\begin{subequations}
    \begin{align}
        &\Dot{t}=\frac{E}{A(r)},\\
        &\Dot{\phi}=\frac{L}{r^2},\\
        &\Dot{r}^2=\frac{1}{B(r)}\lsb\kappa-\frac{E^2}{A(r)}+\frac{L^2}{r^2}\rsb. \label{eqmmmrdot}
    \end{align}
\end{subequations}
Without loss of generality, we have placed the trajectory in the equatorial plane. 
Here, $\kappa=0$ corresponds to null signals and $\kappa=1$ corresponds to timelike signals, respectively. The quantities $E$ and $L$ represent the energy and the angular momentum (per unit mass) of the signal. They can be related to the impact parameter $b$ and the asymptotic velocity $v$ of the trajectory by
\be\label{eqmmmLandE}
E=\frac{1}{\sqrt{1-v^2}},~|L|=\frac{v}{\sqrt{1-v^2}}b.
\ee

\begin{figure}[htp!]
    \centering
\includegraphics[width=0.45\textwidth]{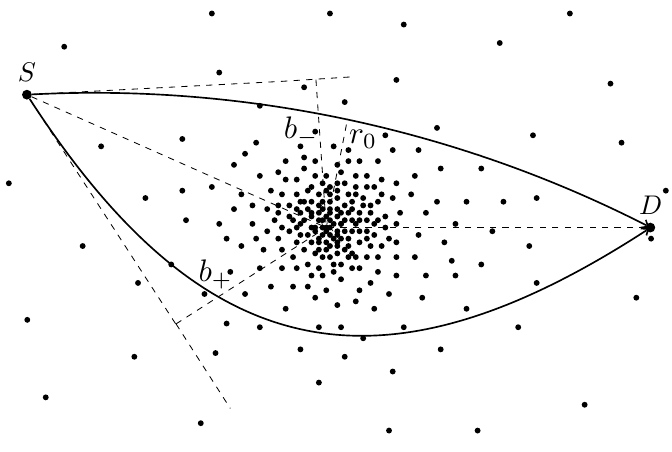}
    \caption{A schematic diagram of the deflection and lensing from source (S) to detector (D) by a density profile. The parameters $b_\pm$ represent the impact parameters of the counterclockwise and clockwise rotating signals. $r_0$ denotes the closest approach of a trajectory. }
    \label{figmmmschdiag}
\end{figure}

The deflection angle $\Delta\phi$ 
of a signal traveling from a source at radius $r_s$ to a detector at radius $r_d$ is then given by 
\begin{align}
\label{eqmmmdeltaphi}
&\Delta\phi=\lsb\int_{r_0}^{r_s}+\int_{r_0}^{r_d}\rsb  \sqrt{\frac{B}{r^2}}\frac{LA}{\sqrt{r^2A\left(E^2-\kappa A\right)-L^2A^2}}\mathrm{d}r,
\end{align}
where $r_0$ is the periapsis radius, defined by $\dd r/\dd \phi|_{r=r_0}$. Using Eq. \eqref{eqmmmrdot}, this definition also provides the following relation between $L$ and $r_0$ by 
\be\label{eqmmmLandr0}
L=r_0\sqrt{\frac{E^2}{A(r_0)}-\kappa},
\ee
which, after applying Eq. \eqref{eqmmmLandE}, can be converted into a relation between $b$ and $r_0$
\begin{align}\label{eqmmmpinfty}
    \frac{1}{b}=\frac{1}{r_0}\sqrt{\frac{E^2-\kappa}{E^2/A(r_0)-\kappa}}\equiv p\lb\frac{1}{r_0}\rb.
\end{align}
Here, in the second step, we introduce a function $p(x)$ that will be used in Sec. \ref{subsecmmmdefangcase1}.

The deflection 
in Eq. \eqref{eqmmmdeltaphi}, however, is usually not integrable to yield a closed-form solution. Therefore, the goal of this section is to find an approximate method to evaluate these integrals,  leveraging the series solutions of the metric obtained in Sec. \ref{secmmmmetricandsol}. 
It turns out that we can apply the same change of variables and expansion of the integrand as employed in \cite{Huang:2020trl,Jia:2020xbc}, although the resulting expansion may differ due to the presence of logarithmic terms in this case. 
Denoting the inverse function of $p(x)$ in Eq. \eqref{eqmmmpinfty} as $q(x)$, which satisfies 
\be\label{eqmmmqinfty}
\frac{1}{r_0}=q\lb\frac{1}{b}\rb,
\ee
the change of variables from $r$ to $u$ is simply given by the relation
\begin{align}
    \frac{1}{r}=q\left(\frac{u}{b}\right). \label{eqmmmqofu}
\end{align}
By using this, it is straightforward to show that Eq. \eqref{eqmmmdeltaphi} becomes
\begin{align}
&\Delta\phi=\lsb\int_{\sin{\beta_s}}^{1}+\int_{\sin{\beta_d}}^{1}\rsb y\lb\frac{u}{b}\rb\frac{\mathrm{d}u}{\sqrt{1-u^2}},\label{eqmmmdphitrans}
\end{align}
where the integrand is given by
\begin{align}
y\lb\frac{u}{b}\rb=\frac{\sqrt{B(1/q)}}{p'(q)q}\frac{u}{b},\label{eqmmmyinftrans}
\end{align}
and $q$ should be interpreted as a function of $u$, as given in Eq. \eqref{eqmmmqofu}.
The $\beta_{s,d}$ in the lower limits are given by 
\be\label{eqmmmbeta}
\beta_{s,d}=\arcsin{\left[ b\cdot p\left(\frac{1}{r_{s,d}}\right)\right]},
\ee
and correspond to the apparent angles of the trajectories at the source and detector \cite{Huang:2020trl}.

The next key step is to approximate the integrand in Eq. \eqref{eqmmmyinftrans} to facilitate the integration. 
To achieve this, we divide the following discussion into two subsections, each corresponding to a different case described in Sec. \ref{secmmmmetricandsol}.

\subsection{Asymptotic case \label{subsecmmmdefangcases}}

\subsubsection{Asymptotic case with integer power}\label{subsecmmmdefangcase1}

For the asymptotic metric solutions presented in Sec. \ref{subsecmmmnont}, if the logarithmic terms are absent, the results for the integrals in Eq. \eqref{eqmmmdeltaphi} have already been obtained in \cite{Huang:2020trl,Jia:2020xbc}. Therefore, in this work, we extend those results to the case where logarithmic terms are present. 

For this case, we simply perform a Taylor expansion of the integrand $y(u/b)$ in Eq. \eqref{eqmmmyinftrans} for small $u$. We then find that
\be
y\lb\frac{u}{b}\rb=\sum_{n=0}^{\infty}\sum_{m=0}^{n}y_{n,m}\lb\frac{u}{b}\rb^n\lb\ln{\frac{u}{b}}\rb^m.
\ee
Here, the coefficients $y_{n,m}$ can be determined by the metric functions $A(r),~B(r)$ or their expansions in Eq. \eqref{eqmmmabinfser}.
Again, for simplicity, we list only the first few of them as 
\begin{widetext}
\begin{subequations}
\label{eqmmmynmfirstfewinf}
\begin{align}
&y_{0,0}=1,\\
&y_{1,0}=\frac{b_{1,0}}{2}+\frac{a_{1,1}-a_{1,0}}{2 v^2},~y_{1,1}=-\frac{b_{1,1}}{2}+\frac{a_{1,1}}{2 v^2},\\
&y_{2,0}=\frac{4 b_{2,0}-b_{1,0}^2}{8} +\frac{- a_{1,0} \left(4a_{1,1}+2 b_{1,0}-b_{1,1}\right)+a_{1,1} b_{1,0}+4 a_{1,0}^2-4 a_{2,0}+2 a_{2,1}}{4 v^2}+\frac{ a_{1,1}^2-2 a_{1,0} a_{1,1}}{4 v^4},\\
&y_{2,1}=\frac{b_{1,0} b_{1,1}-2 b_{2,1}}{4} +\frac{ a_{1,1} \left(b_{1,0}-b_{1,1}\right)+ a_{1,0} \left(b_{1,1}-4 a_{1,1}\right)+2 a_{1,1}^2+2 a_{2,1}}{2 v^2}+\frac{a_{1,1}^2}{2 v^4},\\
&y_{2,2}=\frac{4 b_{2,2}-b_{1,1}^2}{8} -\frac{ a_{1,1} b_{1,1}-2 a_{1,1}^2}{2 v^2}.
\end{align}
\end{subequations}
\end{widetext}

The integrability of $\Delta\phi$ in Eq. \eqref{eqmmmdphitrans} depends on whether we can evaluate integrals of the following form
\be I_{n,m}(\beta_i,b)=\int_{\sin{\beta_i}}^1\lb\frac{u}{b}\rb^n\lb\ln{\frac{u}{b}}\rb^m\frac{\mathrm{d}u}{\sqrt{1-u^2}}
,~~i=s,d. \label{eqmmmiintres} \ee
Fortunately, this is always possible (see Appendix \ref{secmmmappint} for the proof), and the result is approximately a power series in $1/b$ and $\ln (b)$. The expressions for the first few orders are explicitly given in Eq. \eqref{eqmmmfirstfewinm}. By substituting these into Eq. \eqref{eqmmmdphitrans}, we finally obtain the result as
\begin{align}\label{eqmmmdphidtininm}
&\Delta\phi=\sum_{i=s,d}\sum_{n=0}^{\infty}\sum_{m=0}^{n}y_{n,m}I_{n,m}(\beta_i,b).
\end{align}

Eq. \eqref{eqmmmdphidtininm} serves as the master formula, which will be applied in Sec. \ref{secmmmappgl} to determine the deflection for specific matter distributions. By inspecting the form of $I_{n,m}$ (see Eqs. \eqref{eqmmmintegralnandm} and \eqref{eqmmmfirstfewinm}), we observe that the leading order of $I_{n,m}$ is proportional to $(\ln b)^m/b^n$. Consequently, the deflection 
is expressed as a quasi-series in $\ln(b)$ and $b$, with a weak dependence on $r_s$ and $r_d$ through $\beta_i$. This dependence accounts for the finite-distance effect of the source and detector, and it is through this effect that we establish an GL equation and solve for the apparent angles of the lens images in Sec. \ref{secmmmgl}. For this purpose, it is also useful to expand Eq. \eqref{eqmmmdphidtininm} in the small $b/r_{s,d}$ limit. To the first few orders of $\ln(b),~M/b$ and $b/r_{s,d}$, we find
\begin{widetext}
\begin{align}\label{eqmmmdphiexpand}
    \Delta\phi=&\sum_{i=s,d} \frac{\pi}{2}+\frac{1}{b}\left[\frac{b_{1,0}+b_{1,1}}{2}-\frac{a_{1,0}}{2 v^2}+\lb\frac{b_{1,1}}{2}-\frac{a_{1,1}}{2 v^2}\rb\ln{\frac{b}{2}}\right]+\frac{p_{2,0}+p_{2,1}\ln{b}+p_{2,2}\ln^2{b}}{b^2}-\frac{b}{r_i}\nonumber\\
    &-\frac{b}{8r_i^2}\lb2 b_{1,0}+b_{1,1}+\frac{2 a_{1,0}+a_{1,1}}{v^2}\rb-\frac{b\ln{r_i}}{4r_i^2}\lb b_{1,1}+\frac{a_{1,1}}{v^2}\rb+\mathcal{O}\lb\epsilon\rb^3.
\end{align}
Here and henceforth, $\epsilon$ denotes the infinitesimal terms, which could represent $M/b,\, b/r_{s,d}$, or even higher-order terms such as $\ln(b/M)/b,\,\ln (r_{s,d}/M)/r_{s,d}$ and
\begin{subequations}
    \begin{align}
p_{2,0}=&-\frac{\pi }{384}\left[\left(\pi ^2-6+12 \ln^2{2}-12 \ln{2}\right) \left(b_{1,1}^2-4 b_{2,2}\right)+12 \left(b_{1,0}^2-4 b_{2,0}\right)+12 (2 \ln{2}-1) \left(b_{1,0} b_{1,1}-2 b_{2,1}\right)\right]\nonumber\\
&+\frac{\pi}{96v^2}\left\{\left[\pi ^2+12 (\ln{2}-2) \ln{2}\right]\left(2 a_{1,1}^2-a_{1,1} b_{1,1}\right) +12 (\ln{2}-1) \left(4 a_{1,0} a_{1,1}-a_{1,1} b_{1,0}-a_{1,0} b_{1,1}-2 a_{2,1}\right)\right.\nonumber\\
&\left.+24 \left(a_{1,0}^2-a_{2,0}\right)-12 a_{1,0} b_{1,0}\right\}-\frac{\pi}{8v^4}a_{1,1}\lsb a_{1,0}+(\ln{2}-1) a_{1,1}\rsb,\\
p_{2,1}=&-\frac{\pi}{32}\left[2 \left(b_{1,0} b_{1,1}-2 b_{2,1}\right)+(2 \ln{2}-1) \left(b_{1,1}^2-4 b_{2,2}\right)\right]\nn\\
&+\frac{\pi}{8v^2}\left[2 (\ln{2}-1) \left(2 a_{1,1}^2-a_{1,1} b_{1,1}\right)-a_{1,1} b_{1,0}-2 a_{2,1} +a_{1,0} \left(4 a_{1,1}-b_{1,1}\right)\right]-\frac{\pi  a_{1,1}^2}{8 v^4},\\
p_{2,2}=&\frac{\pi}{32}  \left(4 b_{2,2}-b_{1,1}^2\right)+\frac{\pi  \left(2 a_{1,1}^2- a_{1,1} b_{1,1}\right)}{8 v^2}.
    \end{align}
\end{subequations}
In the limit where the source and observer distances approach infinity, the deflection angle simplifies to the following expression, accurate to the first two orders as 
\begin{align}
\Delta\phi=&\pi+\frac{1}{b}\lsb b_{1,0}+\lb1+\ln{\frac{b}{2}}\rb b_{1,1}-\lb a_{1,0}+\ln{\frac{b}{2}}a_{1,1}\rb\frac{1}{v^2}\rsb\nonumber\\
&-\frac{\pi}{192b^2}\left\{K_1 \left(b_{1,1}^2-4 b_{2,2}\right)+12 T_1 \left(b_{1,0} b_{1,1}-2 b_{2,1}\right)+12 \left(b_{1,0}^2-4b_{2,0}\right)+\frac{24}{v^4} a_{1,1} \left(T_2 a_{1,1}+2 a_{1,0}\right)\right.\nonumber\\
&\left.-\frac{4}{v^2}\lsb K_2 \left(2 a_{1,1}^2-a_{1,1} b_{1,1}\right)+6 T_2 \left(-a_{1,1} b_{1,0}-a_{1,0} b_{1,1}+4 a_{1,0} a_{1,1}-2 a_{2,1}\right)+12 \left(2 a_{1,0}^2-a_{1,0} b_{1,0}-2 a_{2,0}\right)\rsb\right\}\nonumber\\
&+\mathcal{O}\lb\frac{\ln^3{b}}{b^3}\rb,
\label{eqmmmdphif2order}
\end{align}
where $T_n=2\ln{2b}-n,~
        K_n=3T_n^2+\pi^2-3n-6~~(n=1,~2)$.
By substituting Eq. \eqref{eqmmmabinffirst},
the deflection in Eq. \eqref{eqmmmdphif2order} can be expressed in terms of the coefficients of the density function as 
\begin{align}
\Delta\phi=&\pi+\frac{2}{b}\lb1+\frac{1}{v^2}\rb\lsb M+4\pi\lb1+\ln{\frac{b}{2}}\rb\rho_3\rsb\nonumber\\
&+\frac{2}{b^2}\left\{\frac{3\pi}{8}Q_1+\frac{\pi^3}{2}(\pi^2-9)\rho_3^2-8\pi\rho_4+\frac{\pi}{v^2}\lsb\frac{3}{2}Q_2+\pi^2\lb2\pi^2-\frac{723}{32}\rb\rho_3^2-\pi\rho_4\rsb\right\}+\mathcal{O}\lb\frac{\ln^3{b}}{b^3}\rb,
\end{align}
\end{widetext}
where
\begin{align}  Q_1=&\lsb2\pi\lb2\ln{2b}-1\rb\rho_3+M\rsb^2,\nonumber\\ Q_2=&\lsb\frac{\pi}{12}\lb48\ln{2b}-15\rb\rho_3+M\rsb^2.
\end{align}
Higher-order results can also be easily obtained, though they are too cumbersome to display here. It becomes more apparent that the $n$-th order of the deflection is the quotient of an $n$-degree polynomial in $\ln b$ and $b^n$. If the coefficients $a_{n,m>1}$ and $b_{n,m>1}$ are set to zero in Eq. \eqref{eqmmmabinfser}, or equivalently if $\rho_3$ is set to zero in Eq. \eqref{eqmmmdensityinfty1}, as in many spacetimes that exhibit only integer-power asymptotic expansions of the metric without logarithmic terms, then the $\ln b$ coefficients in the deflection vanish, and the result aligns with previous findings \cite{Huang:2020trl}. 

\subsubsection{Asymptotic case with non-integer power \label{subsubsecmmmdefangcase2}}

For this case, we perform a Taylor expansion of the integrand $y(u/b)$ for small $u/b$, and find that
\begin{align}
    y\lb\frac{u}{b}\rb=\sum_{n=0}^{\infty}\sum_{m=0}^{t-1}y_{n,m}\lb\frac{u}{b}\rb^{n+m\delta},
\end{align}
where $\delta=s/t$. The coefficients $y_{n,m}$ can be determined by the metric functions $A(r)$ and $B(r)$, or by their expansions in Eqs. \eqref{eqmmmnonintegerA} and \eqref{eqmmmnonintegerB}. Since $t$ is not fixed, the upper limit of $m$ is also not fixed. Therefore, we list only the first few coefficients below as
\begin{subequations}\label{eqmmmnonintegerys}
\begin{align}
    y_{0,0}=&1,\\
    y_{0,1}=&\frac{4 \pi  \rho _2}{\delta -1}\lb1-\frac{1}{v^2}\rb,\\
    y_{0,2}=&\frac{24 \pi ^2 \rho _2^2}{\left(\delta -1\right)^2}+\frac{8 \pi ^2 \left(7 \delta ^2+13 \delta +4\right) \rho _2^2}{\left(\delta -1\right)^2 \delta  \left(\delta +1\right) v^2}\nonumber\\
    &+\frac{32 \pi ^2 \rho _2^2}{\left(\delta -1\right) \delta  v^4}.
\end{align}
\end{subequations}

The integrability of $\Delta\phi$ in Eq. \eqref{eqmmmdphitrans} depends on whether the integrals defined in Eq. \eqref{eqmmmnonintegarintegrate} can be evaluated.
\begin{align}
    I_{n,m}(\beta_i)=&\int_{\sin{\beta_i}}^1\frac{u^{n+m\delta}}{\sqrt{1-u^2}}\mathrm{d}u\nonumber\\
    =&\int_{\beta_i}^{\frac{\pi}{2}}\left(\sin{\theta}\right)^{n+m\delta}\mathrm{d}\theta,\quad i=s,d. \label{eqmmmnonintegarintegrate}
\end{align}
This integral can be expressed using the hypergeometric function, as given in Eq. \eqref{eqmmmnmdintres} in Appendix \ref{secmmmappint}.
By substituting these results into Eq. \eqref{eqmmmdphitrans}, we obtain the final results as shown in 
\begin{align}
    \Delta\phi=\sum_{i=s,d}\sum_{n=0}^{\infty}\sum_{m=0}^{t-1}\frac{y_{n,m}}{b^{n+m\delta}}I_{n,m}(\beta_i).
\end{align}
In the limit of infinite source and observer distances, and to leading order, the deflection angle is given by
\begin{align}
    \Delta\phi=\pi+\frac{4 \pi  \rho _2}{\delta -1}\lb1-\frac{1}{v^2}\rb\frac{\sqrt{\pi } \Gamma \left(\frac{\delta+1}{2}\right)}{ \Gamma \left(\frac{\delta}{2}+1\right)}\frac{1}{b^{\delta}}.
\end{align}

\subsection{Finite boundary case} \label{subsecmmmdeffb}

When the density profile has a finite radius $R$, the deflection must be calculated using the metrics in Eq. \eqref{eqmmmarbrsol2}. In this case, for most gravitational lensing scenarios, the source and detector are located outside the radius $R$, while the signal traverses through the matter inside $R$. Consequently, the corresponding integral for the deflection angle in Eq. \eqref{eqmmmdeltaphi} needs to be divided into four segments: from the source to radius $R$, from $R$ to the minimum radius $r_0$, from $r_0$ back to $R$, and finally from $R$ to the detector. The first and last segments are described by the Schwarzschild metric with a mass $M$. By evaluating the integrals for these two segments, we obtain the corresponding deflection as given in \cite{Jia:2020xbc,Liu:2015zou}
\begin{align}
    \Delta\phi_{\mathrm{Sch}}=&   \sum_{i=s,d}\left\{\frac{M^2}{4R^2v^4}\left[2\tan{x}\left(\sec^2{x}-6v^2\right)\right.\right.\nn\\&\left.\left.-3v^2\left(v^2+4\right)\left(\cot{x}-x\csc^2{x}\right)\right]\right.\nn\\&\left.-\frac{MR}{r_i^2}\frac{\left(v^2-1\right)\sin{x}}{2v^2}+x\right.\nn\\&\left.-\frac{R\sin{x}}{r_i}+\frac{M}{R}\tan{\left(\frac{x}{2}\right)}\frac{v^2-\sec{x}}{v^2}\right\}+\mathcal{O}\lb \epsilon\rb^3,
    \label{eqmmmschintres}
\end{align}
where $x=\arcsin\left(b/R\right)$.

For the segments inside $R$, we should use the metric in Eq. \eqref{eqmmmarbrsol2} and expand the integrand $y(u/b)$ around $u=b/b_R$, where $b_R$ is the impact parameter when $r_0=R$. According to Eq. \eqref{eqmmmpinfty}, we obtain 
\begin{align} \label{eqmmmbR}
b_R=&R\sqrt{\frac{E^2/A(R)-\kappa}{E^2-\kappa}}\\
=&R\sqrt{1+\frac{2M}{\lb R-2M\rb v^2}}.\label{eqmmmbRsim}
\end{align}
This expansion is effectively equivalent to expanding the integrand at $r=R$.
The result of this expansion is
\be 
y\left(\frac{u}{b}\right)=\sum_{n=0}^{\infty}y_n\left(\frac{u}{b}-\frac{1}{b_R}\right)^n,
\label{eqmmmfiniteboundaryexpandy}
\ee
where $y_n$ are the coefficients determined by the metric in Eq. \eqref{eqmmmarbrsol2}. The value for the first coefficient is
\be
y_0=\frac{2 a_0 \sqrt{b_0} \left(a_0 \left(v^2-1\right)+1\right)}{2 a_0^2 \left(v^2-1\right)+2 a_0-a_1 R},
\label{ycoe0}
\ee
where $a_{0,1}$ and $b_0$ are given in Eqs. \eqref{eqmmmabcoeffR}.
Although higher-order coefficients can also be determined, they are not presented here due to their length. The deflection within the density bulge can then be computed by substituting Eq. \eqref{eqmmmfiniteboundaryexpandy} into Eq. \eqref{eqmmmdphitrans} and integrating from $\beta_R$ to 1. 
Here, $\beta_R$ is given by Eq. \eqref{eqmmmbeta} with $r_{s,d}$ replaced by $R$, i.e., 
\be\label{eqmmmbetaR}
\beta_R=\arcsin{\lsb b\cdot p\lb\frac{1}{R}\rb\rsb}.
\ee
The integrability of this integral depends on whether we can evaluate the following 
\begin{align}\label{eqmmmindef}
I_n(\beta_R)=\int_{\sin\beta_R}^1\frac{\left(u/b-1/b_R\right)^n}{\sqrt{1-u^2}}\mathrm{d}u
\end{align}
to obtain a closed form. Fortunately, the evaluation is relatively straightforward, and the results are expressed in terms of elementary trigonometric functions, as detailed in Appendix \ref{secmmmappint}. 

Finally, by substituting $I_n$ and $y_n$ from Eq. \eqref{eqmmmfiniteboundaryexpandy} into Eq. \eqref{eqmmmdphitrans}, and incorporating Eq. \eqref{eqmmmschintres}, we obtain the total deflection for this case in the form of 
\begin{align}
\Delta \phi=&\Delta\phi_{\mathrm{Sch}}+2\sum_{n=0}^\infty y_nI_n(\beta_R). 
\label{eqmmmdeffiniteg}
\end{align}

\section{Application to known matter densities and Gravitational lensing\label{secmmmappgl}}

In this section, we apply the results from the previous section to several well-known matter distributions. Our focus will be on: (1) models commonly used in gravitational lensing studies, such as the SIS and PIS models \cite{kochanek1991implications,narayan1988gravitational,congdon2018principles,Kochanek:1994vw,kent1988model,Kochanek:1999rj,Rusin:2002tq}; (2) models that are significant in astronomy for representing the matter distributions of galaxies or galaxy clusters, such as the NFW or gNFW models; and (3) models with mathematical advantages, such as the uniform distribution model, which allows for an exact solution to the metrics and facilitates direct comparison with our results \cite{Wald:1984rg}. Considering these aspects, a particularly suitable choice is the general density model with three free parameters \cite{Hernquist:1990be,Zhao:1995cp,Weber:2009pt,Kravtsov:1997dp,zhao1996analytical,kazantzidis2004generating} given by 
\begin{align}
\rho(r)=\frac{\rho_c}{\left(r/r_m\right)^{\gamma}\left[1+\left(r/r_m\right)^{\alpha}\right]^{\left(\beta-\gamma\right)/\alpha}}. 
\label{eqmmmmmbf}
\end{align}
Here, $\rho_c$ represents the overall density scale, and $r_m$ denotes the size of the main galaxy (or cluster) halo, the indices $\alpha\geq 0,~\beta\geq 2$ and $\gamma\geq 0$ control the shape around $r_m$, the slope at infinity, and the cusp index at small $r$, respectively. 
This density function naturally encompasses all the aforementioned cases, as well as many other models. A brief summary of the sub-model names and their corresponding parameter choices is provided in Tab. \ref{tab:bfd}. This model and its sub-models have also been extensively used to fit observational data \cite{Kravtsov:1997dp,Knudson:2001fp,Keeton:1997by,Mortlock:2000zu,Donaldson:2021byu,kulessa1992mass}. 

\begin{table}[htp!]
    \centering
    \scalebox{1.0}{
    \begin{tabular}{l|c|c}
    \hline\hline
    Model name & $\alpha, \beta, \gamma$ &  Ref.\\
    \hline\hline
    gNFW& $1, 3, \gamma$  &  \cite{Chae:2013zha,Newman:2012nv,Courteau:2013cjm,Baxter:2017csy} \\
        ~~NFW & $1, 3 ,1$ &  \cite{Navarro:1996gj,Chemin:2011mf,Chae:2013zha,Newman:2012nv,Courteau:2013cjm,Kravtsov:2012zs,Pratt:2019cnf,Merten:2014wna,Navarro:1995iw,lokas2001properties,Lilleengen:2022xcw} \\
        \hline
        Power-law & $\slashi{\alpha}, \gamma, \gamma$  & \cite{Chae:2013zha} \\
        ~~SIS & $\slashi{\alpha}, 2, 2$ &  \cite{Remmen:2021tyj}\\
        \hline
        Hernquist & $1, 4, 1$ &   \cite{Hernquist:1990be,bollati2023dynamical} \\
        \hline
        ~~PIS & $2,2,0$ & \cite{begeman1991extended,yang2024black,Chemin:2011mf,Randriamampandry:2014eoa,burkert1995structure} \\
        \hline\hline
    \end{tabular}}
    \caption{The general three-parameter model Eq. \eqref{eqmmmmmbf} encompasses several density models. Among these, the NFW model is a subclass of the gNFW model, while the SIS model is a subclass of the power-law model. The notation $\slashi{\alpha}$ indicates that the parameters $\alpha$ are excluded from this particular model.}
    \label{tab:bfd}
\end{table}

The structure of this section follows that of the previous one. In Subsec. \ref{subsecmmmappa}, we study distributions that extend to infinity, whereas in Subsec. \ref{subsecmmmappb}, we examine distributions truncated at a finite radius. 

\subsection{Densities extended to infinity\label{subsecmmmappa}}

For the density profile given by Eq. \eqref{eqmmmmmbf}, when $\beta$ is greater than 2, the distribution extends to infinity without causing asymptotic non-flatness. By expanding this density profile at infinity, we can straightforwardly obtain the series in the form of
\begin{align}
\rho(r)=\sum_{n=0}\frac{\rho^\prime_n}{r^{\beta+n\alpha}},~~(\alpha>0),\label{eqmmmghrhoexp}
\end{align}
with the coefficients given as
\begin{align}
\label{eqmmmrhopres}
\rho^\prime_n=\frac{(-1)^n\rho_c r_m^{\beta+n\alpha}}{n!}\lb \frac{\beta-\gamma}{\alpha}\rb_n,
\end{align}
where $(x)_n$ denotes the Pochhammer symbol of $x$. In the limit $\alpha\to 0$, the distribution described by Eq. \eqref{eqmmmmmbf} effectively reduces to a power-law function 
\begin{align}
\rho(r)=\frac{\rho_c r_m^\beta}{r^\beta},
\end{align}
which corresponds to the power-law model in Tab. \ref{tab:bfd}, with $\beta$ taking the role of $\gamma$. Therefore, this case will not be treated separately here. 

When, and only when, both $\alpha$ and $\beta>2$ are integers, can these expansions be substituted into the procedure outlined in Sec. \ref{sssecmmmdensitywithip} to compute $m(r)$ and $P(r)$, as well as the metric functions $A(r),~B(r)$, using Eqs. \eqref{eqmmmpressureinfty1} to \eqref{eqmmmabinffirst}. This indicates that for all models in Tab. \ref{tab:bfd}, except for the power-law model with non-integer $\beta=\gamma$, and the SIS and PIS models, which are asymptotically non-flat, the method and results from Sec. \ref{sssecmmmdensitywithip} are applicable. Accordingly, we will first analyze the gNFW, NFW, and Hernquist models, before turning to the power-law model with $\gamma\geq 3$ or potentially fractional $\gamma$. In the next subsection, we will focus on the SIS and PIS models truncated at a finite radius. 

\subsubsection{The gNFW and NFW models}

For simplicity, we list only the first few coefficients of the metric functions $A(r)$ and $B(r)$ for the gNFW model, which includes the NFW model as a special case and has become increasingly popular in recent years for modeling galaxy (or cluster) matter distributions. Due to the presence of a nonzero $\rho_3$, the metric functions for this model inherently contain logarithmic terms. 

By substituting the indices $(\alpha,~\beta,~\gamma)=(1,~3,~\gamma)$ into Eq. \eqref{eqmmmghrhoexp}, we obtain the series expansion for $\rho(r)$ in the gNFW model. Furthermore, by substituting this expression into Eqs. \eqref{eqmmmmrinfinrho} and \eqref{eqmmmabinffirst}, we derive the corresponding series expansions for the mass function $m(r)$ and the metric functions $A(r)$ and $B(r)$. Below, we explicitly present the first few coefficients of the metric functions as 
\begin{subequations}
\label{eqmmmabninfint}
    \begin{align}
a_{1,0}=&4M_0  \left(C_{\gamma}-1+\ln{r_m}\right),\\
a_{1,1}=&-4M_0 ,\\
a_{2,0}=&\frac{M_0 }{4} \lsb M_0  \left(-28 C_{\gamma }-28 \ln{r_m}+13\right)+8 \left(\gamma -3\right) r_m\rsb,\\
a_{2,1}=&7M_0 ^2,\\
b_{1,0}=&-4M_0  \left(C_{\gamma}+\ln{r_m}\right),\\
b_{1,1}=&4M_0 ,\\
b_{2,0}=&4M_0  \lsb4M_0  \left(C_{\gamma }+\ln{r_m}\right)^2-\left(\gamma -3\right) r_m\rsb,\\
b_{2,1}=&-32M_0 ^2 \left(C_{\gamma}+\ln{r_m}\right),\\
b_{2,2}=&16M_0 ^2,    
\end{align}
\end{subequations}
where we have defined $M_0 =2\pi r_m^3\rho_c$ and
\begin{align}
    C_{\gamma}=\begin{cases}
\displaystyle  \sum_{i=1}^{2-\gamma}\frac{1}{i}, &\quad\gamma=0,~1,~2 ,\\
\displaystyle   \frac{\Gamma'\left(3-\gamma\right)}{\Gamma\left(3-\gamma\right)}+c_E, &\quad \gamma=\text{otherwise}.
    \end{cases}
\end{align}
Here, $c_E$ represents the Euler–Mascheroni constant. The specific results for the NFW model can be obtained by setting $\gamma=1$ in the expressions derived above.

\begin{figure}[htp!]
    \centering
\includegraphics[width=0.45\textwidth]{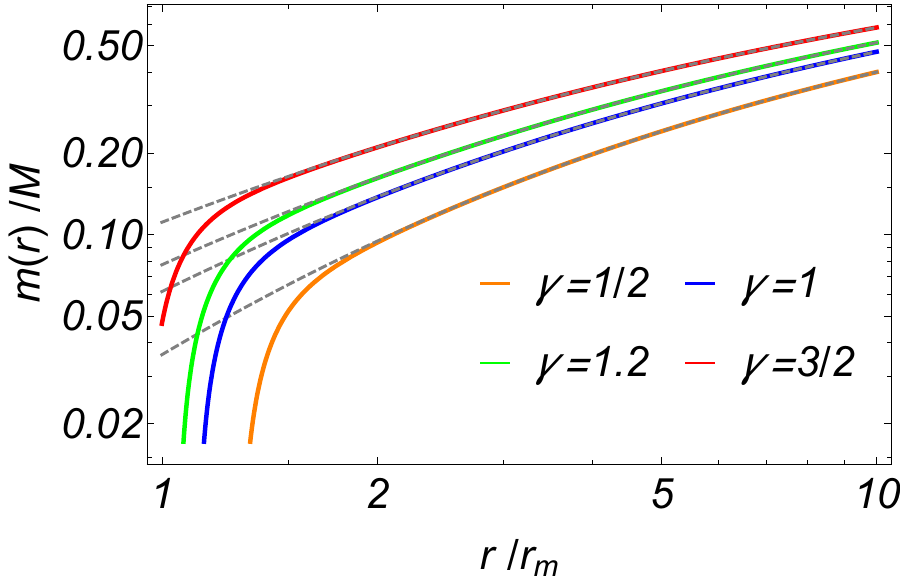}\\
(a)\\
\includegraphics[width=0.45\textwidth]{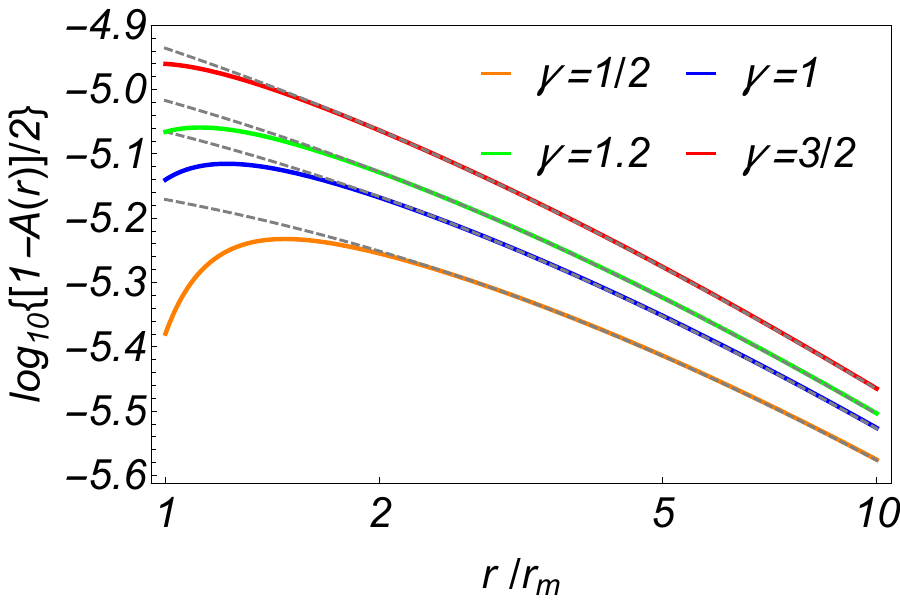}\\
(b)\\
\includegraphics[width=0.45\textwidth]{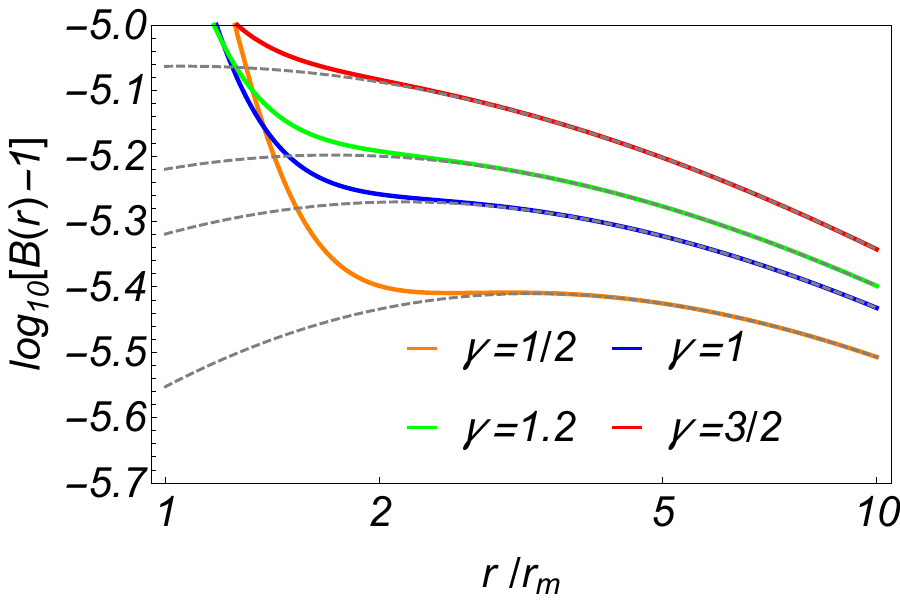}\\
(c)
\caption{\label{figmmmgnfwnfwplot} Mass and metric functions in the gNFW model. (a) $m(r)$, obtained perturbatively from Eq. \eqref{eqmmmmrinfinrho} (solid lines) compared to the exact solution from Eq. \eqref{eqmmmmrgnfwsol} (dashed lines). (b) Gravitational potential $-\Phi(r)$ (or $[1-A(r)]/2$) and (c) the metric function $B(r)-1$, both obtained perturbatively using Eq. \eqref{eqmmmabninfint} (solid lines) and numerically (dashed lines). Other parameters chosen are $\rho_c=4\times10^8~M_{\odot}/\mathrm{kpc}^3$ and $r_m=18~\mathrm{kpc}$ \cite{VERITAS:2010meb,Lilleengen:2022xcw}.}  
\end{figure}

To validate the accuracy of these series results, we compare them with the corresponding functions derived from the numerical integration of the TOV equations, as illustrated in Fig. \ref{figmmmgnfwnfwplot}. 
For $m(r)$, the gNFW density permits symbolic integration, which can be expressed in terms of the incomplete Beta-function
\begin{align}
    m(r)=-4 \pi  (-1)^{\gamma } \rho_c r_m^3 Beta\lb-\frac{r}{r_m}; 3-\gamma ,\gamma -2\rb
    \label{eqmmmmrgnfwsol},
\end{align}
It can be simplified in the NFW limit to 
\begin{align}
    m(r)=4 \pi  \rho_c r_m^3 \lsb-\frac{r}{r+r_m}+\ln{\lb1+\frac{r}{r_m}\rb}\rsb.
\end{align}
We directly compare the series result for $m(r)$ with its corresponding symbolic solutions. For the potential $\Phi(r)$ and the metrics functions $A(r)$ and $B(r)$, only numerical results are available for comparison with the series solutions. Since the potential $\Phi(r)$ is small ($\Phi(r)\ll 1$), we approximate $[1-A(r)]/2=[1-\rme^{2\Phi(r)}]/2\approx -\Phi(r)$, Thus, $-\Phi(r)$ and $[1-A(r)]/2$ are displayed in the same plot. 
The series solutions used in Fig. \ref{figmmmgnfwnfwplot} are truncated to order 8 for $m(r),~\Phi(r),~A(r)$, and to order 9 for $B(r)$. 

It is observed that, for all choices of $\gamma=\{0.5,\,1.0,\,1.2,\,1.5\}$, the series results match the symbolic and numerical solutions extremely well when $r\gtrsim \mathcal{O}(r_m)$, but become inaccurate when $r\lesssim\mathcal{O}(r_m)$. This behavior is expected, as the series solution is an asymptotic approximation. The scale $\mathcal{O}(r_m)$ represents the minimal radius at which the series solution remains valid for studying trajectory behavior, including the bending angle. Dimensional analysis of Eq. \eqref{eqmmmabninfint} shows that the fundamental convergent scale of the series for $m(r),~A(r)$ and $B(r)$ is determined by $r_m$ in the density profile given by Eq. \eqref{eqmmmghrhoexp}. This convergent scale remains valid even in the case of power-law models discussed in Subsec. \ref{sssecmmmpwlg2}. From Fig. \ref{figmmmgnfwnfwplot}, we observe that the metric functions approach $1$ as the radius goes to infinity, indicating that the spacetime becomes flat at infinity.

By substituting the
coefficients from Eq. \eqref{eqmmmabninfint} into Eq. \eqref{eqmmmdphiexpand}, we obtain the deflection angle in the gNFW model for different values of $\gamma$ as 
\begin{align}\label{eqmmmgNFWexpand}
&\Delta\phi_{\mathrm{gNFW}}\nn\\
=&\sum_{i=s,d}\frac{\pi}{2}+\frac{2r_m}{b}\frac{M_0 }{r_m}\lb1+\ln{\frac{b}{2r_m}}-C_{\gamma}\rb\lb1+\frac{1}{v^2}\rb\nonumber\\
    &+\lb\frac{r_m}{b}\rb^2\frac{M_0 }{r_m}\lb z_{2,0}+z_{2,1}\ln{\frac{b}{r_m}}+z_{2,2}\ln^2{\frac{b}{r_m}}\rb\nonumber\\
    &-\frac{b}{r_i}-\frac{r_m b}{2r_i^2}\frac{M_0 }{r_m}\left[1-2\lb C_{\gamma}-\ln{\frac{b}{r_m}}\rb\right.\nonumber\\
    &\left.+\frac{1}{v^2}\lb2C_{\gamma}-2\ln{\frac{b}{r_m}}-3\rb\right]\nonumber\\
    &-\lb1-\frac{1}{v^2}\rb\frac{r_m b}{r_i^2}\frac{M_0 }{r_m}\ln{\frac{r_i}{b}}+\mathcal{O}\lb \epsilon^3\rb,
\end{align}
where the coefficients $z_{n,m}$ in this case are given by
\begin{subequations}
    \begin{align}
        z_{2,0}=&\frac{3\pi}{8}\frac{M_0 }{r_m}\left[\lb2C_{\gamma}-2\ln{2}+1\rb^2+\frac{\pi^2}{3}-3\right]\nonumber\\
        &+\frac{\pi\lb3-\gamma\rb}{2}+\frac{\pi}{16v^2}\left\{\frac{3}{8}\frac{M_0 }{r_m}\left[\lb16C_{\gamma}-16\ln{2}+5\rb^2\right.\right.\nonumber\\
        &\left.\left.+\frac{64}{3}\pi^2 -241\right]+8\lb3-\gamma\rb\right\}+\frac{2\pi}{v^4}\frac{M_0 }{r_m}\lb C_{\gamma}-\ln{2}\rb,\\
        z_{2,1}=&-\frac{M_0 }{r_m}\left[\frac{3\pi}{2}\lb2C_{\gamma}-2\ln{2}+1\rb\right.\nonumber\\
        &\left.+\frac{\pi}{v^2}\left(\frac{15}{4}+12 C_{\gamma}-12\ln{2}\right)+\frac{2\pi}{v^4}\right],\\
        z_{2,2}=&\frac{3\pi}{2}\frac{M_0 }{r_m}\lb1+\frac{4}{v^2}\rb.
\end{align}
\end{subequations}

\begin{figure}[htp!]
    \centering
\includegraphics[width=0.45\textwidth]{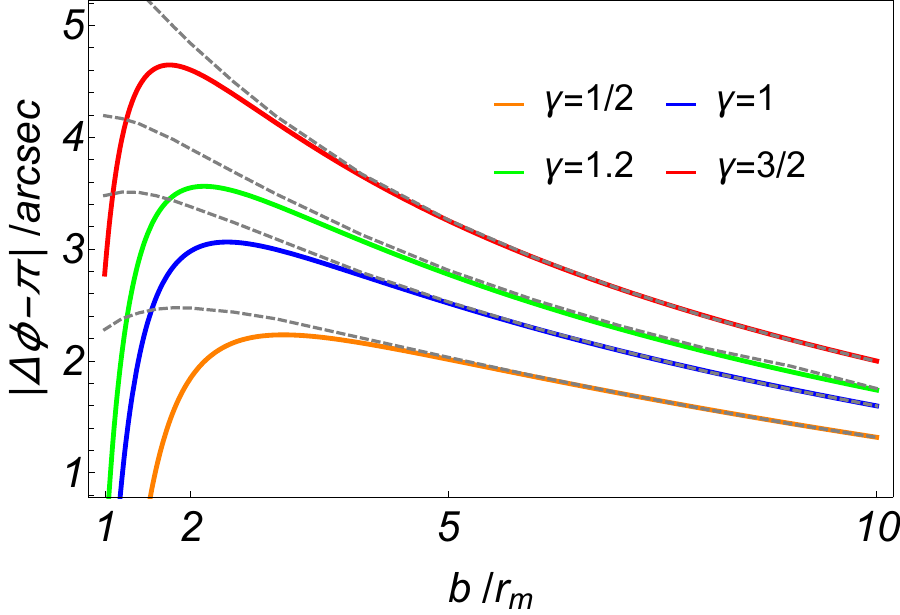}
\caption{\label{figmmmgnfwdef} Deflection angle obtained from Eq. \eqref{eqmmmgNFWexpand} (solid lines) in the gNFW model for different values of $\gamma$. The parameters are chosen as $\rho_c=4\times 10^8~M_{\odot}/\mathrm{kpc}^3$, $r_m=18~\mathrm{kpc}$, $r_{s}=r_d=10^7 r_m$. For these values, $M_0$ is approximately $6.2\times 10^{-6} r_m$. }  
\end{figure}

According to Eq. \eqref{eqmmmgNFWexpand}, it is clear that the expanded $\Delta\phi$ is a series in terms of $r_m/b$ and $b/r_i$. The coefficients of this series are functions of the velocity $v$ and quantity $M/r_m=2\pi\rho_cr_m^2$. In Fig. \ref{figmmmgnfwdef}, we plot the dependence of $\Delta\phi$ on the impact parameter $b$ for the gNFW model with $\gamma=\{0.5,\,1.0,\,1.2,\,1.5\}$. It is observed that similar to the results for $m(r),\, A(r)$ and $B(r)$, the series results closely match the numerical values (dashed lines) when $b\gtrsim r_m$, and decrease monotonically as $b$ increases. 

\subsubsection{The Hernquist model}

For the Hernquist model, by substituting the indices $(\alpha,~\beta,~\gamma)=(1,~4,~1)$ into the density expression given by Eq. \eqref{eqmmmmmbf} and Eq. \eqref{eqmmmghrhoexp}, and then using Eqs. \eqref{eqmmmmrinfinrho} and \eqref{eqmmmabinffirst}, we can obtain the series expansions for the mass function $m(r)$, and the metric functions $A(r)$ and $B(r)$. Here, we present the first few coefficients of these metric functions as 
\begin{subequations}
\label{eqmmmhmser}
    \begin{align}
        &a_{1,0}=-2M_0 ,\\
        &a_{2,0}=2M_0 r_m,\\
        &a_{3,0}=2 M_0  r_m \left(8 M_0 -15 r_m\right),\\
        &a_{4,0}=\frac{M_0  r_m \left(33 M_0 ^2-95 M_0  r_m+60 r_m^2\right)}{30},\\
        &b_{1,0}=2M_0 ,\\
        &b_{2,0}=4 M_0  \left(M_0 -r_m\right),\\
        &b_{3,0}=2 M_0  \left(4 M_0 ^2-8 M_0  r_m+3 r_m^2\right),\\
        &b_{4,0}=8 M_0  \left(2 M_0 ^3-6 M_0 ^2 r_m+5 M_0  r_m^2-r_m^3\right).
    \end{align}
\end{subequations}
As can be seen from Eq. \eqref{eqmmmmrhqmodel}, here $M_0 =2\pi\rho_c r_m^3$ represents the total mass at infinite radius. Note that all $a_{n,m>0}$ and $b_{n,m>0}$ vanish because $\rho_3=0$ in the Hernquist model. Consequently, this model does not exhibit logarithmic terms in its asymptotic expansion. 

By substituting the coefficients of these metric functions into Eq. \eqref{eqmmmdphiexpand}, we can obtain the deflection $\Delta\phi_{\mathrm{H}}$ for the Hernquist model in the form of 
\begin{align}
    \Delta\phi_{\mathrm{H}}=&\sum_{i=s,d}\frac{\pi}{2}
    +\frac{r_m}{b}\frac{M_0}{r_m}\lb1+\frac{1}{v^2}\rb\nonumber\\
    &+\lb\frac{r_m}{b}\rb^2\frac{M_0}{r_m}\left[\frac{\pi}{8}\lb\frac{3M_0}{r_m}-4\rb-\frac{\pi}{2v^2}\lb1-\frac{3M_0}{r_m}\rb\right]\nonumber\\
    &+\lb\frac{r_m}{b}\rb^3\frac{M_0}{r_m}\left\{\frac{5}{3}\lb\frac{M_0}{r_m}\rb^2-\frac{4M_0}{r_m}+2\right.\nonumber\\
    &+\frac{1}{v^2}\lsb15\lb\frac{M_0}{r_m}\rb^2-\frac{226}{15}\frac{M_0}{r_m}+2\rsb\nonumber\\
    &\left.-\frac{1}{v^4}\frac{M_0}{r_m}\lb2-\frac{5M_0}{r_m}\rb-\frac{1}{3v^6}\lb\frac{M_0}{r_m}\rb^2\right\}\nonumber\\
    &-\frac{b}{r_i}-\frac{r_m b}{2r_i^2}\frac{M_0}{r_m}\lb1-\frac{1}{v^2}\rb-\frac{b^3}{6r_i^3}\nonumber\\
    &-\frac{r_m^2 b}{6r_i^3}\frac{M_0}{r_m}\left[\frac{3M_0}{r_m}-4+\frac{2}{v^2}\lb1-\frac{3M_0}{r_m}\rb+\frac{3}{v^4}\frac{M_0}{r_m}\right]\nonumber\\&+\mathcal{O}\lb\epsilon^4\rb.
    \label{eqmmmdphihm}
\end{align}
Similar to the deflection angle in Eq. \eqref{eqmmmgNFWexpand} for the gNFW model, Eq. \eqref{eqmmmdphihm} is expressed as a series in terms of $r_m /b$ and $b/r_i$, with coefficients that are rational functions of $v^2$ and the quantity $M_0/r_m=2\pi\rho_c r_m^2$.

\begin{figure}[htp!]
    \centering
\includegraphics[width=0.45\textwidth]{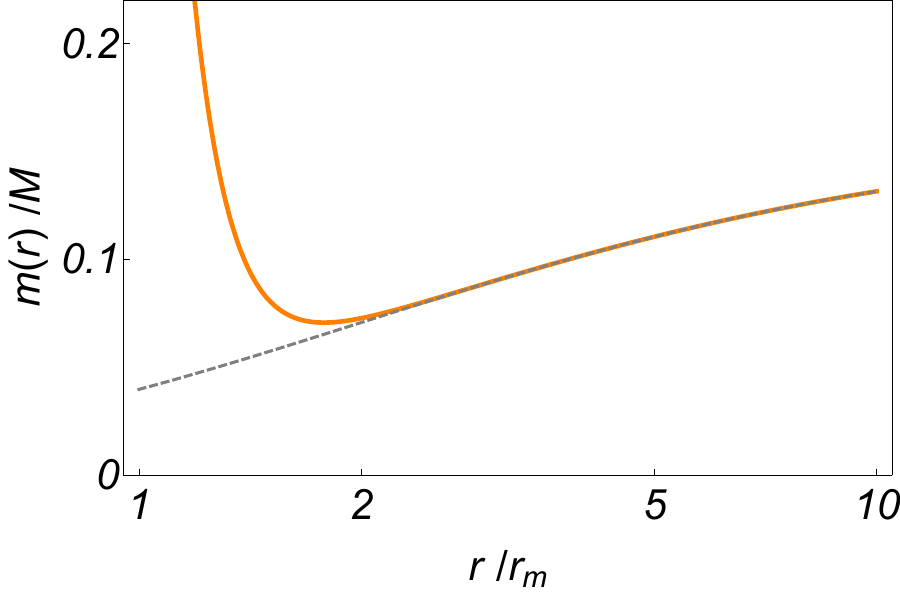}\\
(a)\\
\includegraphics[width=0.45\textwidth]{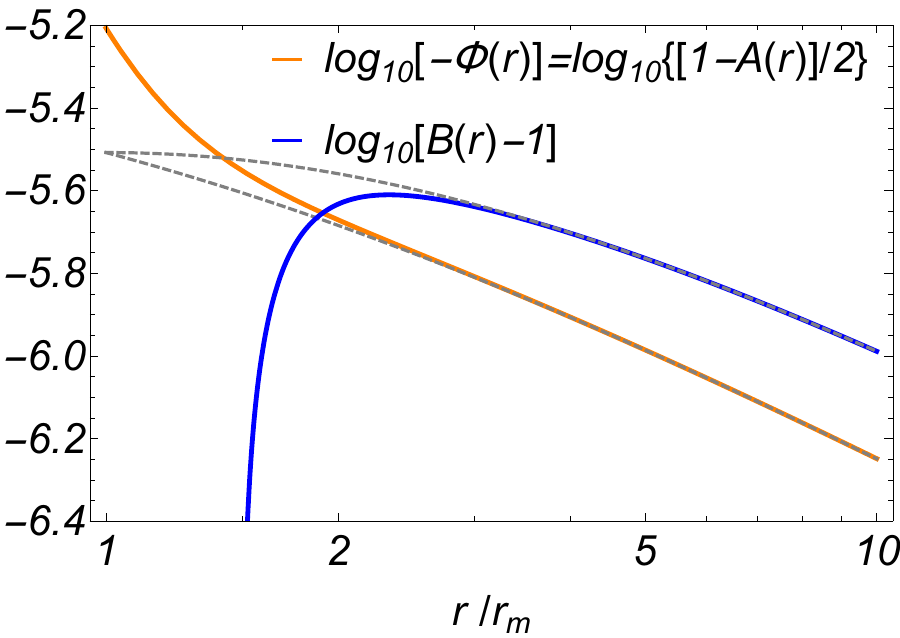}\\
(b)\\
\includegraphics[width=0.45\textwidth]{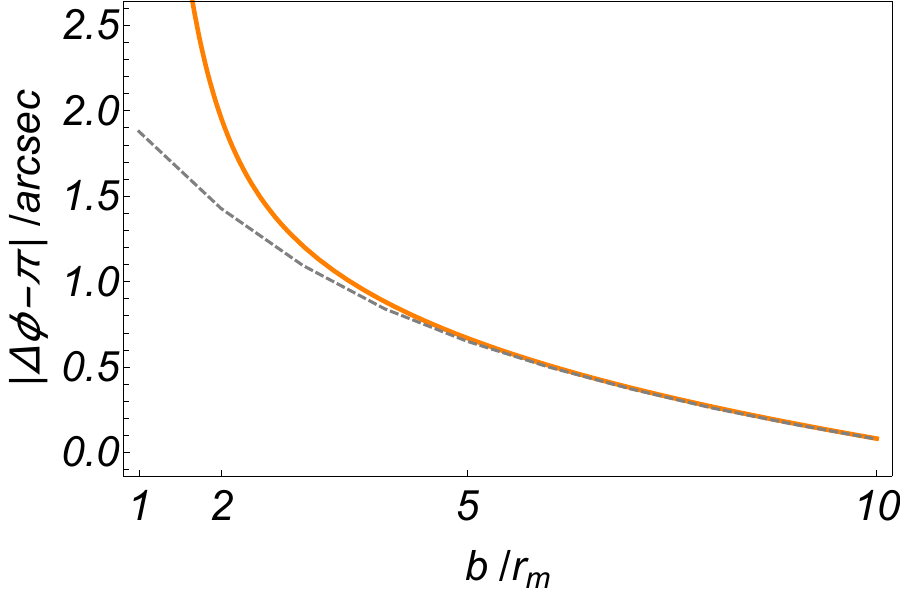}\\
(c)
\caption{\label{figmmmhernquistplot}The Hernquist model results. 
(a) Mass function $m(r)$: the perturbative expression from Eq. \eqref{eqmmmmrinfinrho} (solid line) and the exact solution from Eq. \eqref{eqmmmmrhqmodel} (dashed line). (b) Potential $\Phi(r)$ (or equivalently $1-A(r)$) and the metric function $B(r)-1$. (c) Deflection angle $\Delta\phi$ obtained perturbatively (solid line) and numerically (dashed line). Parameters chosen are $\rho_c=4\times10^8~M_{\odot}/\mathrm{kpc}^3$, $r_m=18~\mathrm{kpc}$ \cite{bollati2023dynamical} and $r_s=r_d=10^7 r_m$. The $m(r),~\Phi(r),~A(r)$ are to 8th order, $B(r),~\Delta\phi$ are to 9th and the 3rd order, respectively.  }
\end{figure}

To verify the correctness of the series for the mass function $m(r)$, the metric functions with coefficients given in Eq. \eqref{eqmmmhmser}, and the deflection angle $\Delta\phi_\mathrm{H}$ from Eq. \eqref{eqmmmdphihm}, we compare them with the corresponding functions obtained through numerical integration, as shown in Fig. \ref{figmmmhernquistplot}. For $m(r)$ in the Hernquist model, the density allows for a straightforward analytical solution
\begin{align}
    m(r)=\frac{M_0 r^2}{\left(r+r_m\right)^2}.
    \label{eqmmmmrhqmodel}
\end{align}
It is observed that the series solutions for both the mass function and the metric functions agree very well with the numerical solutions when $r$ is larger than approximately $\mathcal{O}(r_m)$. Furthermore, in this case, the mass function converges to the constant value $M$, which contrasts with the behavior observed in Fig. \ref{figmmmgnfwnfwplot}. For the deflection angle, Eq. \eqref{eqmmmdphihm} indicates that the series converges to a value where $b\approx \mathcal{O}(r_m)$, since $r_m\gg M$. Consequently, as long as the series is computed to a sufficiently high order, the series deflection angle should match the numerical value within this range of $b$. This agreement is precisely what is observed in Fig. \ref{figmmmhernquistplot} (c). 

\subsubsection{power-law model with \texorpdfstring{$\gamma>2$}{gamma>2}\label{sssecmmmpwlg2}}

For the power-law density $\rho(r)=\rho_c r_m^\gamma/r^\gamma$, a straightforward integration yields the following mass
\begin{align}
    m(r)=\int_0^r4\pi x^2\frac{\rho_c r_m^{\gamma}}{x^{\gamma}}\mathrm{d}x+C_m.
\end{align}
If $\gamma\leq 2$, the mass function $m(r)$ diverges as $r\to\infty$, and the metric function $B(r)$ and $A(r)$ also blow up. If $2< \gamma\leq 3$, the metric functions will converge, although the mass function will still grow to infinity. For $\gamma>3$, a divergence occurs as $r\to 0$ due to the strong singularity in the density profile at the center. However, in such cases, the density profile is often modified near the center to resolve this issue.
In this subsection, we assume that $\gamma>2$ and that a density modification is applied at $r=0$, if necessary, to ensure that the mass remains finite as $r\to 0$. We will then focus on the asymptotic behavior of the mass and metric functions under these assumptions. The mass can be obtained from the indefinite integral of the density so that
\begin{align}
    m(r)=m_0+\begin{cases}
        \displaystyle\frac{4\pi\rho_c r_m^{\gamma}}{3-\gamma}r^{3-\gamma},~&\gamma>2,~\gamma\neq 3,\\
        \displaystyle 4\pi\rho_c r_m^3\ln{\frac{r}{2\pi\rho_c r_m^3}},~&\gamma=3.
    \end{cases}
\label{eqmmmmasspowerlaw}
\end{align}
Here, if $\gamma>3$, $m_0$ represents the total mass. Otherwise, $m_0$ is just a constant, and Eq. \eqref{eqmmmmasspowerlaw} is valid only outside the region where the density has been modified.

By using the above mass function, the metric functions are determined from Eq. \eqref{eqmmmabinfser} as follows. For $\gamma=3$,
\begin{subequations}
\begin{align}
A(r)=&1+\frac{-2 \left(4 \pi  \rho _c r_m^3+C_m\right)-8 \pi  \rho _c r_m^3\ln{r}}{r}\nonumber\\
&+\frac{\pi  \rho _c r_m^3 \left(13 \pi  \rho _c r_m^3+7 C_m\right)+28 \pi ^2 \rho _c^2 r_m^6\ln{r}}{r^2}\nonumber\\&+\mathcal{O}(r)^{-3},\label{eqmmmg3a}\\
B(r)=&1+\frac{2C_m+8 \pi  \rho_c r_m^3\ln{r}}{r}\nonumber\\
&+\frac{4C_m^2+32 \pi  C_m \rho_c r_m^3\ln{r}+64 \pi ^2 \rho _c^2 r_m^6\ln^2{r}}{r^2}\nonumber\\&+\mathcal{O}(r)^{-3}\label{eqmmmg3b},
\end{align}
\end{subequations}
where $C_m=m_0-4\pi\rho_c r_m^3\ln{\left(2\pi\rho_c r_m^3\right)}$. By substituting the expansion coefficients of these metric functions into Eq. \eqref{eqmmmdphiexpand}, the deflection in the power-law model is given by  
\begin{align}
\Delta\phi_{\gamma=3}=&\sum_{i=s,d}\frac{\pi}{2}+\frac{M_0}{b}\left(2+\frac{m_0}{M_0}+2\ln{\frac{b}{2M_0}}\right)\lb1+\frac{1}{v^2}\rb\nonumber\\
    &+\lb\frac{M_0}{b}\rb^2\lb z_{2,0}+z_{2,1}\ln{\frac{2b}{M_0}}+z_{2,2}\ln^2{\frac{2b}{M_0}}\rb\nonumber\\
    &-\frac{b}{r_i}-\frac{b M_0}{r_i^2}\left(\frac{1+m_0/M_0}{2}-\frac{3+m_0/M_0}{2v^2}\right)\nonumber\\
    &-\frac{b M_0}{r_i^2}\lb1-\frac{1}{v^2}\rb\ln{\frac{r_i}{M_0}}+\mathcal{O}(\epsilon^3) \label{eqmmmg3d},
\end{align}
where $M_0=2\pi\rho_c r_m^3$ and 
\begin{subequations}
    \begin{align}
    z_{2,0}=&\frac{\pi}{16}\lsb2\pi^2\lb1+\frac{4}{v^2}\rb-3\lb4+\frac{27}{v^2}\rb\rsb\nonumber\\&-\frac{\pi}{8}\frac{m_0}{M_0}\lb6+\frac{15}{v^2}+\frac{8}{v^4}\rb\nonumber\\&+\frac{3\pi}{8}\frac{m_0^2}{M_0^2}\lb1+\frac{4}{v^2}\rb,\\
    z_{2,1}=&\frac{3\pi}{2}\frac{m_0}{M_0}\lb1+\frac{4}{v^2}\rb-\frac{\pi}{4}\lb6+\frac{15}{v^2}+\frac{8}{v^4}\rb,\\
    z_{2,2}=&\frac{3\pi}{2}\lb 1+\frac{4}{v^2}\rb.
\end{align}
\end{subequations}
When $\gamma$ is an integer greater than or equal to 4, Eq. \eqref{eqmmmabinfser} ensures that 
$a_{n,m>0}=b_{n,m>0}=0$. Consequently, the metric functions take the form shown in 
\begin{subequations}
\begin{align}
    &A(r)=1-\frac{2m_0}{r}+\frac{\left(\gamma-4\right)!}{\left(\gamma-2\right)!}\frac{8\pi\rho_c r_m^{\gamma}}{r^{\gamma-2}}+\mathcal{O}\lb\frac{1}{r}\rb^{\gamma-1},\label{eqmmmg4a}\\
&B(r)=1+\sum_{n=1}^{\infty}\frac{b_{n,0}}{r^n}\label{eqmmmg4b},
\end{align}
\end{subequations}
where the coefficients are
\begin{align}
    b_{n,0}=\begin{cases}
        \left(2m_0\right)^n, ~~~~~~~~~~~~~~~~~~~~~~~~~~~~~n=1,\cdots,\gamma-3,\\
        2m_0 b_{n-1,0}+\frac{8\pi\rho_c r_m^{\gamma}}{3-\gamma}b_{n-\gamma+2,0}, ~~n=\gamma-2,\cdots.
    \end{cases}
\end{align}
The deflection angle in this case can be obtained from Eq. \eqref{eqmmmdphiexpand} as
\begin{align}
    \Delta\phi_{\gamma_\mathrm{int}}=&\sum_{i=s,d}\frac{\pi}{2}+\frac{1}{b}\lb\frac{b_{1,0}}{2}-\frac{a_{1,0}}{2 v^2}\rb+\frac{p_{2,0}}{b^2}\nonumber\\
    &-\frac{b}{r_i}-\frac{b}{8r_i^2}\lb2 b_{1,0}+\frac{2 a_{1,0}}{v^2}\rb+\mathcal{O}\lb\epsilon\rb^3, \label{eqmmmg4d}
\end{align}
where
\begin{align}
p_{2,0}=&-\frac{\pi}{32}\lb b_{1,0}^2-4b_{2,0}\rb\nonumber\\
&+\frac{\pi}{96 v^2}\lsb24\lb a_{1,0}^2-a_{2,0}\rb-12 a_{1,0} b_{1,0}\rsb.
\end{align}

For $\gamma$ that is not an integer but takes the form $\gamma=k+s/t$, where $k>2$ is an integer and $s/t\neq0$ is a reduced fraction, we substitute $\rho_k=\rho_c r_m^{\gamma}$ and $\rho_{n\neq k}=0$ into Eqs. \eqref{eqmmmnonintegerP} and  \eqref{eqmmmnonintegerab}. This substitution enables us to determine the coefficients of $A(r),~B(r)$, and the deflection angle $\Delta\phi_\gamma$. However, the general symbolic expression for these coefficients depends on $n$ in a highly intricate manner and is therefore omitted here. Instead, we present the results of the mass function $m(r)$ for the specific case of $\gamma=2+\frac{1}{2}$ as an example. By substituting $\gamma=2+\frac{1}{2}$ into Eq. \eqref{eqmmmmasspowerlaw}, we obtain 
\begin{align}
    &m(r)=m_0+8\pi\rho_c r_m^\frac{5}{2}r^\frac{1}{2}.
\end{align}
We also present the results for the metric functions and the deflection angle in the case where $\gamma=2+\frac{1}{2}$.
\begin{subequations}
\begin{align} &A(r)=\sum_{n=0}^{\infty}\sum_{m=0}^{1}a_{n,m}\frac{1}{r^{n+\frac{m}{2}}},\label{eqmmmnonintegerAp}\\
&B(r)=\sum_{n=0}^{\infty}\sum_{m=0}^{1}b_{n,m}\frac{1}{r^{n+\frac{m}{2}}},\label{eqmmmnonintegerBp}
\end{align}
\end{subequations}
where the first few orders of coefficients are
\begin{subequations}\label{eqmmmnonintegerser}
\begin{align}
    &a_{0,0}=1,\\
    &a_{0,1}=-32\pi \rho_2,\\
    &a_{1,0}=-2m_0+\frac{704}{3}\pi^2\rho_2^2,\\
    &a_{1,1}=\frac{32}{63}\pi\lb39m_0\rho_2+416\pi_2^3+21\rho_3\rb,\\
    &b_{0,0}=1,\\
    &b_{0,1}=16\pi\rho_2,\\
    &b_{1,0}=2\lb m_0+128\pi^2\rho_2^2\rb,\\
    &b_{1,1}=16\pi\lb 4m_0\rho_2+256\pi^2\rho_2^3-\rho_3\rb.
\end{align}
\end{subequations}

\begin{figure}[htp!]
    \centering
\includegraphics[width=0.45\textwidth]{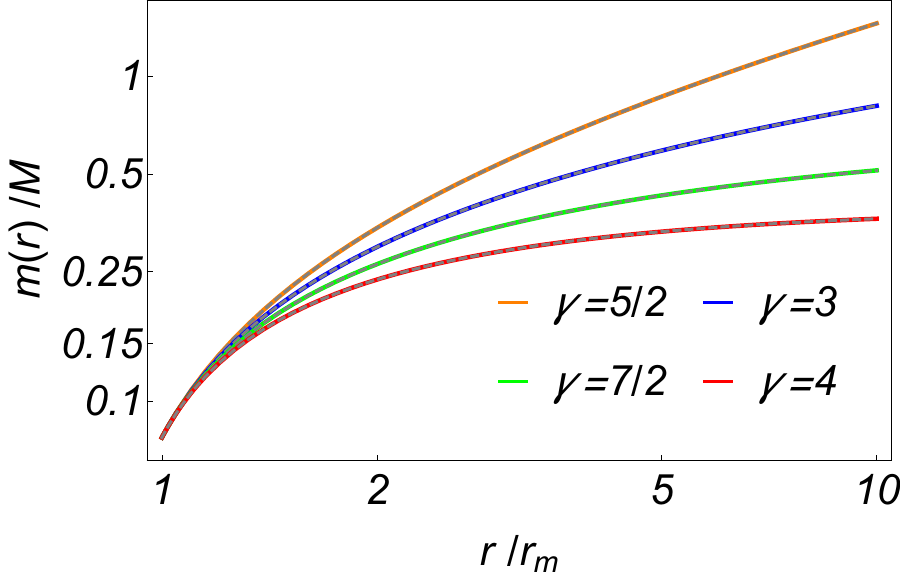}\\
(a)\\
\includegraphics[width=0.45\textwidth]{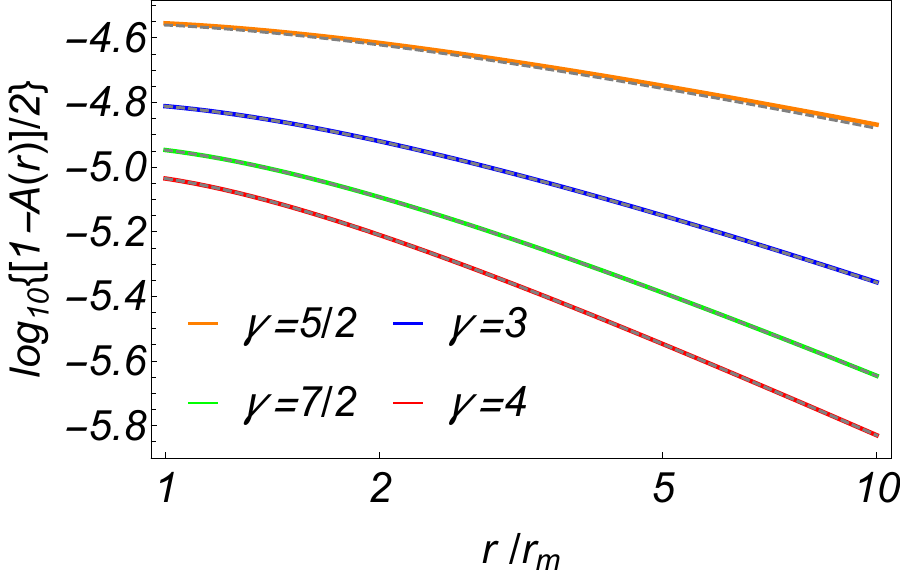}\\
(b)\\
\includegraphics[width=0.45\textwidth]{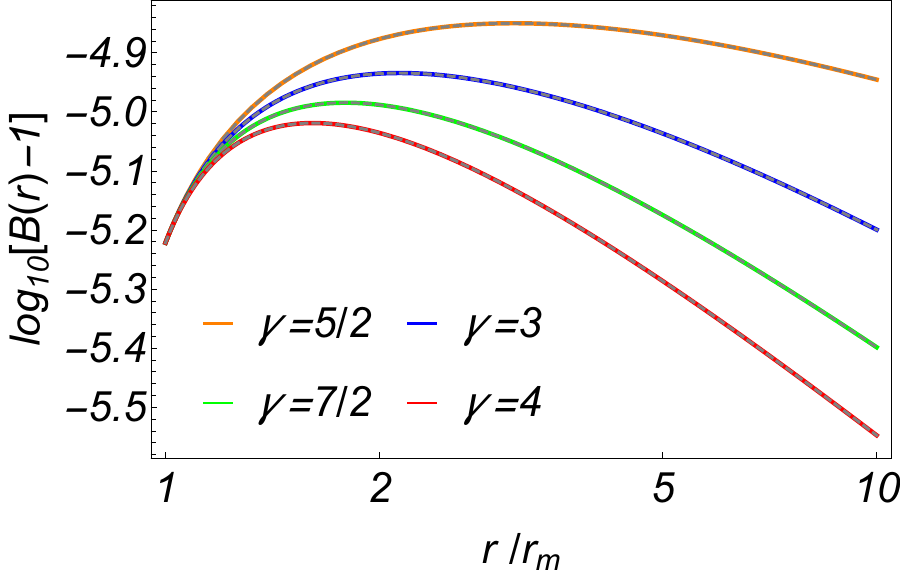}\\
(c)\\
\caption{\label{figmmmpwl1f} Mass and metric functions in power-law models for several $\gamma\geq 3$. 
(a) $m(r)$ from Eq. \eqref{eqmmmmasspowerlaw} (solid lines) and corresponding numerical values (dashed lines). (b) and (c) are the functions $-\Phi(r),~1-A(r)$ and $B(r)-1$ from Eqs. \eqref{eqmmmg3a}-\eqref{eqmmmg3b}, \eqref{eqmmmg4a}-\eqref{eqmmmg4b}, \eqref{eqmmmnonintegerAp}-\eqref{eqmmmnonintegerBp} (solid lines) respectively, and corresponding numerical values (dashed lines). Parameters chosen are $\rho_c=4\times10^8~M_{\odot}/\mathrm{kpc}^3$, $r_m=18~\mathrm{kpc}$ and $r_s=r_d=10^7 r_m$. $m(r),~\Phi(r),~A(r)$ are truncated to 8th order, while $B(r)$ is to 9th order. }
\end{figure}

By substituting Eq. \eqref{eqmmmnonintegerser} into Eq. \eqref{eqmmmnonintegerys}, and combining it with the solution of the integral in Eq. \eqref{eqmmmnonintegarintegrate}, the deflection angle can be determined by
\begin{align}
    \Delta\phi=&\sum_{i=s,d}\frac{\pi}{2}+\frac{1}{b^{\frac{1}{2}}}\frac{4\pi^\frac{3}{2}\rho_2\Gamma{(\frac{3}{4})}}{v^2\Gamma{(\frac{5}{4})}}\lb1+\frac{1}{v^2}\rb\nonumber\\&+\frac{1}{b}\lsb m_0\lb1+\frac{1}{v^2}\rb+\frac{32\pi^2\rho_2^2\lb9v^4+49v^2-12\rb}{3v^4}\rsb\nonumber\\&-\frac{b}{r_i}+\frac{b}{r_i^\frac{3}{2}}\frac{16\pi\rho_2}{81}\lb 73-\frac{27}{v^2}\rb\nonumber\\&+\frac{b}{r_i^2}\left[\frac{m_0}{162}\lb 119-\frac{81}{v^2}\rb\right.\nonumber\\&\left.+\frac{16\pi^2\rho_2^2\lb1639 v^4-14121 v^2+8748\rb}{2187 v^4}\right]\nn\\
&+\mathcal{O}(\epsilon)^{3/2}.  \label{eqmmmnonintegerdp}
\end{align}

\begin{figure}[htp!]
    \centering
    \includegraphics[width=0.45\textwidth]{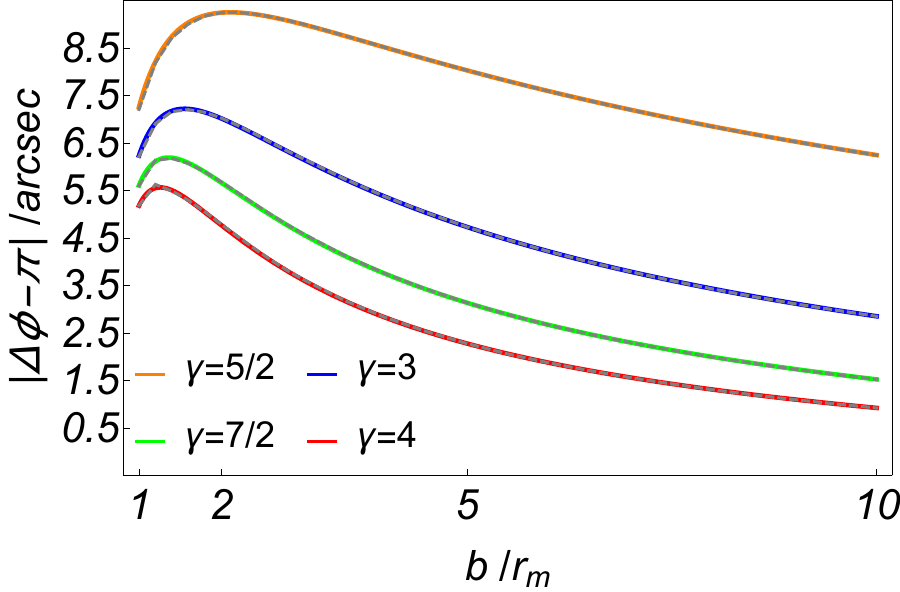}
    \caption{Deflection angle truncated to the 3rd order in power-law models for several $\gamma \geq 2$ from Eqs.~\eqref{eqmmmg3d}, \eqref{eqmmmg4d} and \eqref{eqmmmnonintegerdp} (solid lines), and their numerical solutions (dashed lines), respectively. Parameters chosen are $\rho_c=4\times10^8~M_{\odot}/\mathrm{kpc}^3$, $r_m=18~\mathrm{kpc}$ and $r_s=r_d=10^7 r_m$. }
    \label{figmmmpwlifdf}
\end{figure}

In Fig. \ref{figmmmpwl1f} and Fig. \ref{figmmmpwlifdf}, we present the mass and metric functions, as well as the deflection angles, for the power-law models with four representative values of $\gamma$, namely $\gamma=2\frac{1}{2},\,3,\,3\frac{1}{2},\,4$. The solid lines represent the results obtained using the analytical expressions derived in this subsection, while the dashed lines correspond to the results from numerical integration. 
Similar to the Hernquist model, these functions converge to their true values when $r\gtrsim r_m$. 
For the deflection angle, the series solutions consistently agree with the numerical results. Notably, if $b$ is too large, the finite-distance effects of the source and detector can surpass the leading-order deflection (see Eq. \eqref{eqmmmg3d} and Eq. \eqref{eqmmmnonintegerdp}), causing the deflection to become negative and increase as $b$ grows. This behavior is similar to that observed in the gNFW model. 

\subsection{Density with boundary \label{subsecmmmappb}}

As noted in the previous subsection, the mass function diverges as $r\to\infty$ in certain models, such as the PIS model or the power-law model with $\gamma\leq 3$. For the power-law model with $\gamma<2$, even the metric functions $A(r),~B(r)$ diverge. To satisfy the requirement of asymptotic flatness, these density profiles are typically truncated at a finite radius in practical applications.
In the following, we analyze the deflection in such models, assuming the density is truncated at a radius $R$.

\subsubsection{PIS model}

The PIS model is widely applied in gravitational lensing and dark matter studies \cite{congdon2018principles,kent1988model,yang2024black}. Its density profile within the nonzero range takes the form of
\begin{align}
    \rho_{\mathrm{PIS}}(r)=\frac{\rho_c r_m^2}{r_m^2+r^2},~r\leq R.\label{eqmmmpisdensity}
\end{align}
Using the procedure outlined in Sec. \ref{subsecmmmmetricR}, the metric functions can be derived by expanding the density function around $r=R$ as given in Eq. \eqref{eqmmmdensityR}, and substituting the resulting coefficients into Eq. \eqref{eqmmmmcoeffR} and Eq. \eqref{eqmmmabcoeffR}.

By following this procedure, the mass function within the boundary can be expressed as
\begin{align}
    m(r)=&M+\frac{4\pi R^2r_m^2\rho_c\lb r-R\rb}{R^2+r_m^2}\nonumber\\&+\frac{4\pi Rr_m^4\rho_c\lb r-R\rb^2}{\lb R^2+r_m^2\rb^2}+\mathcal{O}(r-R)^3, \label{eqmmmmrpis}
\end{align}
while the corresponding metric functions inside $R$ are
\begin{subequations}
\label{eqmmmmetricpis}
\begin{align}
A(r)=&1-\frac{2 M}{R}+\lb\frac{2 M}{R^2}+8\pi RP_0\rb(r-R)\nonumber\\
&+\left[\frac{16\pi^2R^3P_0^2+4\pi P_0\lb R-M\rb}{R-2M}-\frac{2M}{R^3}\right.\nonumber\\&\left.+\frac{4\pi r_m^2\rho_c\lb R-M\rb+16\pi^2 r_m^2\rho_c P_0}{\lb R-2M\rb\lb R^2+r_m^2\rb}\right]\lb r-R\rb^2\nonumber\\&+\mathcal{O}(r-R)^3,\label{eqmmmpisa}\\
B(r)=&\lb1-\frac{2 M}{R}\rb^{-1}+\frac{8\pi  R^3  r_m^2\rho_c-2M\lb R^2+r_m^2\rb}{\lb R-2 M\rb^2\lb R^2+r_m^2\rb}\lb r-R \rb\nonumber\\
&-\left[\frac{64\pi^2R^5r_m^4\rho_c^2}{\lb R^2+r_m^2\rb^2\lb2M-R\rb^3}+\frac{8\pi R^4r_m^2\rho_c}{\lb R^2+r_m^2\rb^2\lb2M-R\rb^2}\right.\nonumber\\&\left.-\frac{32M\pi R^2r_m^2\rho_c}{\lb R^2+r_m^2\rb\lb2M-R\rb^3}+\frac{2M}{\lb2M-R\rb^3}\right]\lb r-R\rb^2\nonumber\\&+\mathcal{O}(r-R)^3,\label{eqmmmpisb}\\
C(r)=&R^2+2R\left(r-R\right)+\left(r-R\right)^2. 
\end{align}
\end{subequations}
For this specific case, the mass function can be solved analytically as shown in 
\begin{align}
    m(r)=4\pi r_m^2\rho_c\left[ r-r_m\arctan\left(\frac{r}{r_m}\right)\right],~r\leq R.
\end{align}
The quantity $M=m(R)$ in Eqs. \eqref{eqmmmmrpis} and \eqref{eqmmmmetricpis} represents the total mass within the radius $R$. The parameter $P_0$ denotes the pressure at the boundary $r=R$. This pressure is not set to zero by default, as certain primordial models allow for a nonzero value. In deriving these functions, the condition $m(r=0)=0$ and the matching condition $B(r=R)=1/\left( 1-M/R\right)$ were applied. The metric functions outside the boundary, as previously noted, coincide with those of the Schwarzschild spacetime. 

By substituting the coefficients of the metric functions $A(r)$ and $B(r)$ into Eq. \eqref{eqmmmdeffiniteg}, and setting $P_0=0$ to ensure that $P(R)=0$ at the surface of the sphere, the deflection angle for this case can be determined. Given that $M/R$ is small and $b$ and $R$ are of comparable magnitude, the deflection angle can be expanded into a series in terms of $M/R$, yielding
\begin{widetext}
    \begin{align}
       \Delta\phi=&\Delta\phi_{\mathrm{Sch}}+2\arccos{\frac{b}{R}}+\frac{4\pi R^2r_m^2\rho_c\lb v^2+1\rb\lb b\pi-2\sqrt{R^2-b^2}-2b\arcsin{\frac{b}{R}}\rb}{bv^2\lb R^2+r_m^2\rb}\nonumber\\&+\frac{M}{R}\frac{1}{bR^2v^4\lb R^2+r_m^2\rb}\left\{ -8\pi R^4r_m^2\rho_c\sqrt{R^2-b^2}\lb v^2+1\rb\right.\nonumber\\&\left.+2 R^2\lsb4\pi R^2 r_m^2\rho_c\lb 3v^4+8v^2+1\rb-v^2\lb R^2+r_m^2\rb\lb v^2+1\rb\rsb\lb\sqrt{R^2-b^2}-b\arccos{\frac{b}{R}}\rb\right.\nonumber\\&\left.+2b R^2 v^2\lb R^2+r_m^2\rb\lsb \frac{b}{\sqrt{R^2-b^2}}+\lb v^2+1\rb\arccos{\frac{b}{R}}\rsb\right\}+\mathcal{O}\lb\epsilon^2\rb,
       \label{eqmmmdefpis}
   \end{align} 
\end{widetext}
where $\Delta\phi_{\mathrm{Sch}}$ is the amount of bending outside the sphere, as given in Eq. \eqref{eqmmmschintres}. 

\begin{figure}[htp!]
    \centering
\includegraphics[width=0.45\textwidth]{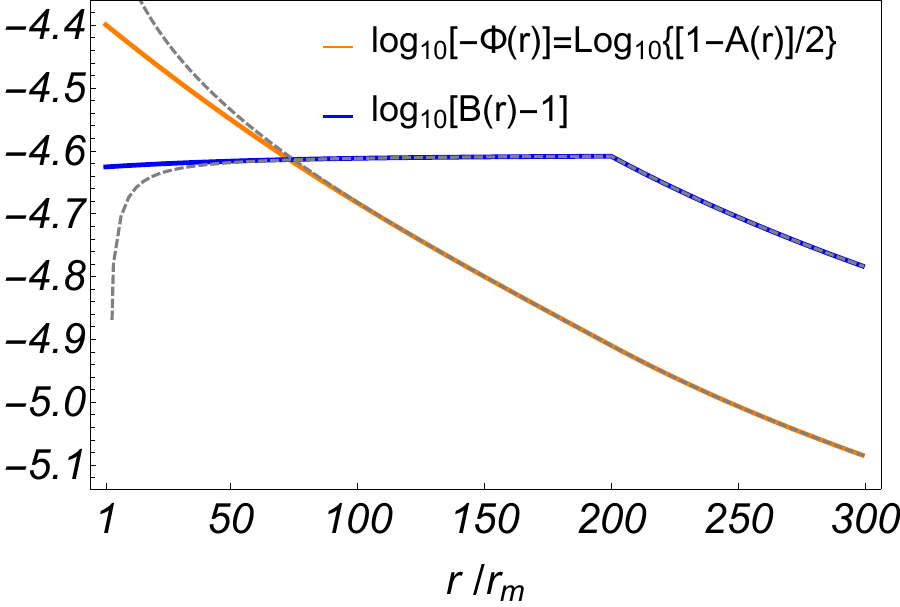}\\
(a)\\
\includegraphics[width=0.45\textwidth]{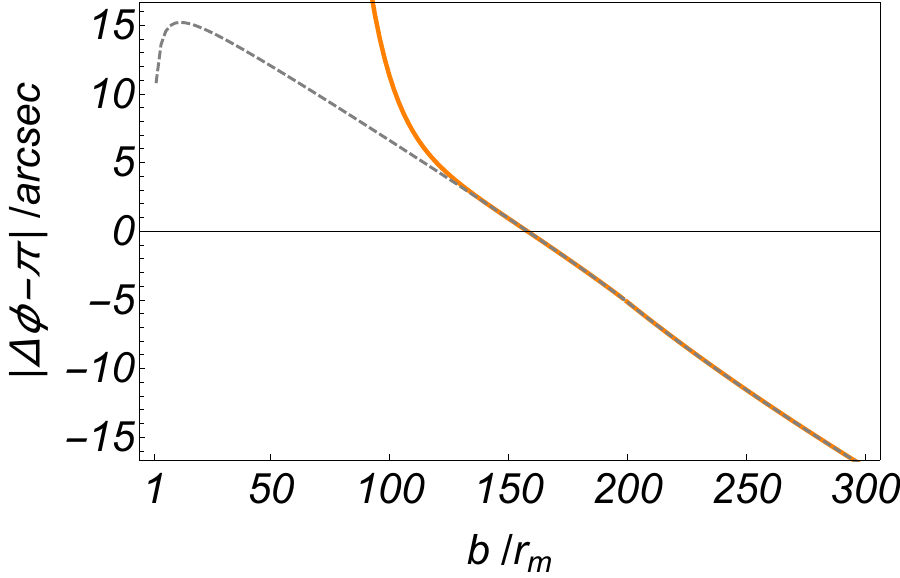}\\
(b)
\caption{\label{figmmmPISplot}  
Perturbative (dashed line) and numerical (solid line) solutions of the metric functions and deflection angle in the PIS model. Parameter values are as follows: (a) $\rho_c=4\times10^{8}~M_{\odot}/\mathrm{kpc}^3$, $r_m=0.18~\mathrm{kpc}$ and $R=200 r_m$. (b) $r_m=18~\mathrm{kpc}$, $R=200 r_m$ and $r_s=r_d=5\times10^6 r_m$. }
\end{figure}

Fig. \ref{figmmmPISplot} shows the metric functions and the deflection angle for the PIS model, assuming the density boundary is located around $R=200 r_m$. The non-smooth transitions at the boundary, evident in all plots, are a characteristic feature of such solutions and arise from the first-order matching condition at this point. Our analytical solutions for the metric functions and the deflection closely align with the numerical results (dashed lines) for the impact parameter ranging from approximately $R/2$ to infinity. However, due to the expansion at $r=R$, the method does not allow $b$ to decrease to zero.

\subsubsection{The power-law model with \texorpdfstring{$\gamma\leq 3$}{gamma<=3}}

Similar to the truncated PIS model, the density function in this case must also be truncated for the method to work effectively, particularly for $\gamma<2$, where the metric functions would otherwise diverge. Notably, power-law models with $\gamma\leq 3$ encompass several commonly used cases, including the uniform density model ($\gamma=0$) \cite{Wald:1984rg,bozza2002gravitational,Virbhadra:1999nm,Virbhadra:2002ju} and the SIS model ($\gamma=2$) \cite{congdon2018principles,Remmen:2021tyj}.
Expanding the power-law density $\rho(r)=\rho_c r_m^\gamma/r^\gamma$ around the truncation point $r=R$ and substituting the coefficients into Eq. \eqref{eqmmmmcoeffR}, the mass function $m(r)$ for $r\leq R$ is given by
\begin{align}
    m(r)=&M+4\pi R^{2-\gamma} r_m^\gamma\rho_c\lb r-R\rb\nonumber\\&+2\pi R^{1-\gamma}r_m^\gamma\rho_c\lb\gamma-2\rb\lb r-R\rb^2+\mathcal{O}(r-R)^3,
\end{align}
and substituting the coefficients into Eq. \eqref{eqmmmabcoeffR}, the metric functions are obtained as
\begin{subequations}[t]
\label{eqmmmarbrpowerlaw}
\begin{align}
A(r)=&1-\frac{2M}{R}+\lb\frac{2M}{R^2}+8\pi RP_0\rb\lb r-R\rb\nonumber\\&+\left\{\lsb \lb\frac{r_m}{R}\rb^\gamma\rho_c+P_0\rsb\frac{\lsb4\pi\lb R-M\rb+16\pi^2P_0R^3\rsb}{\lb R-2M\rb}\right.\nonumber\\&\left.-\frac{2M}{R^3}\right\}\lb r-R\rb^2+\mathcal{O}\lb r-R\rb^3,\\
B(r)=&\lb1-\frac{2M}{R}\rb^{-1}+\frac{8\pi r_m^\gamma\rho_c R^{3-\gamma}-2M}{\lb R-2M\rb^2}\lb r-R\rb\nonumber\\&+\frac{1}{\lb 2M-R\rb^3}\left\{4\pi r_m^\gamma\rho_c\lb R^{3-\gamma}\gamma-16\pi r_m^\gamma\rho_c R^{5-2\gamma}\rb\right.\nonumber\\
        &\left.-2M\left[ 1+4\pi r_m^\gamma\rho_c R^{2-\gamma}\lb\gamma-4\rb\right]\right\}\lb r-R\rb^2\nonumber\\&+\mathcal{O}(r-R)^3,\\
C(r)=&R^2+2R(r-R)+(r-R)^2.
\end{align}
\end{subequations}
Here and throughout this subsection,
\begin{align} 
M=m(R)=-4\pi R^3\rho_c\lb r_m/R\rb^{\gamma}/\lb\gamma-3\rb
\end{align}
represents the total mass within radius $R$ for a given $\gamma$ and other parameters. This result is derived using the condition $m(r=0)=0$ and the matching condition $B(R)=1/\left[ 1-m(R)/R\right]$.
For the uniform density case with $\gamma=0$, the metric functions can be solved exactly for $r\leq R$ as \cite{Wald:1984rg}
\begin{subequations}\label{eqmmmuniformexactmetric}
    \begin{align}
        A(r)=&\frac{1}{12}\lb\sqrt{3-8\pi r^2\rho_c}-3\sqrt{3-8\pi R^2\rho_c}\rb^2,\\
        B(r)=&\lb1-\frac{8}{3}\pi r^2 \rho_c\rb^{-1},\\
        C(r)=&r^2.
    \end{align}
\end{subequations}
Similarly, for the SIS density with $\gamma=2$, the exact metric functions are
\begin{subequations}\label{eqmmmsisexactgeneral}
    \begin{align}
        A(r)=&k_2 \lb\frac{r}{R}\rb^{c_1}\frac{2^{c_2}}{\left\{1+\lb\frac{r}{R}\rb^{-\frac{2k_1}{k_2}}-\frac{1-3k_2}{2k_1}\lsb\lb\frac{r}{R}\rb^{-\frac{2k_1}{k_2}}-1\rsb\right\}^{c_2}},\\
        B(r)=&\frac{1}{k_2},\\
        C(r)=&r^2,
    \end{align}
\end{subequations}
where we choose the boundary condition as $P\lb r=R\rb=P_0=0$. Here, $k_1$, $k_2$, $c_1$ and $c_2$ are\begin{align}
&k_1=\sqrt{128\pi^2r_m^4\rho_c^2-24\pi r_m^2\rho_c+1},\\
&k_2=1-8\pi r_m^2\rho_c,\\
&c_1=-\frac{2\lb k_1-k_2\rb\lb k_2-1\rb}{\lb1+2k_1-3k_2\rb k_2},\\
&c_2=-\frac{2\lb k_2-1\rb^2}{\lb1-3k_2\rb^2-4k_1^2}.
\end{align}
If we choose the boundary condition $P(r=R)=P_0=0$, the pressure solution becomes
\begin{align}
P(r)=&\frac{\lb k_2-1\rb^2\lb r^{-\frac{2k_1}{k_2}}-R^{-\frac{2k_1}{k_2}}\rb}{8\pi r^2\lsb r^{-\frac{2k_1}{k_2}}\lb2k_1+3k_2-1\rb+R^{-\frac{2k_1}{k_2}}\lb2k_1-3k_2+1\rb\rsb}.    \label{eqmmmpratbforsis}
\end{align}
The deflection angle can be obtained by substituting the coefficients of the metric functions $A(r)$ and $B(r)$ into Eq. \eqref{eqmmmdeffiniteg}. Expanding this expression for small $M/R$, we obtain
\begin{widetext}
\begin{align}
       \Delta\phi=&\Delta\phi_{\mathrm{Sch}}+2\arccos{\frac{b}{R}}+\frac{8\pi R^{2-\gamma}r_m^{\gamma}\rho_c\left( v^2+1\right)\left( -\sqrt{R^2-b^2}+b\arccos{\frac{b}{R}}\right)}{bv^2}\nonumber\\&+\frac{M}{R}\frac{2}{bv^4}\left[ v^2\left( \frac{R^2}{\sqrt{R^2-b^2}}+v^2\sqrt{R^2-b^2}\right)-4\pi R^{2-\gamma}r_m^{\gamma}\rho_c\sqrt{ R^2-b^2}\left(3v^4+9v^2+2\right)\right.\nonumber\\&\left.+4b\pi R^{2-\gamma}r_m^{\gamma}\rho_c\left( 3v^4+8v^2+1\right)\arccos{\frac{b}{R}}\right]+\mathcal{O}\left(\epsilon^2\right).
       \label{eqmmmdefpower}
   \end{align} 
\end{widetext}
The deflection angles for the uniform and SIS density cases can be obtained by substituting $\gamma=0$ and $\gamma=2$ into this equation, respectively. 
The deflection angle simplifies in the limit of an infinitely distant source and detector, as well as in the null signal limit, by setting $r_{s,d}\to\infty$ and substituting $v=1$ into Eq. \eqref{eqmmmdefpower}, respectively.
\begin{figure}[htp!]
    \centering
\includegraphics[width=0.9\linewidth]{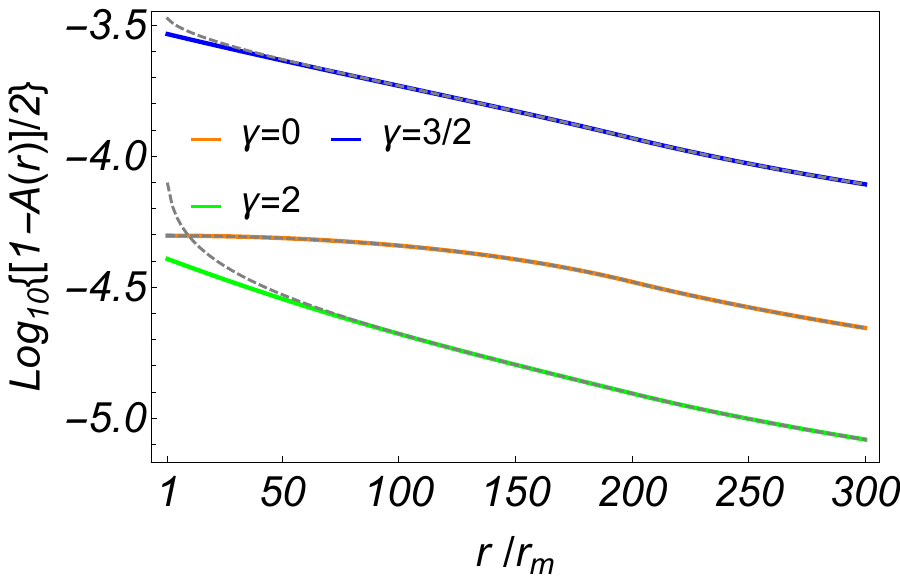}\\
(a)\\
\includegraphics[width=0.9\linewidth]{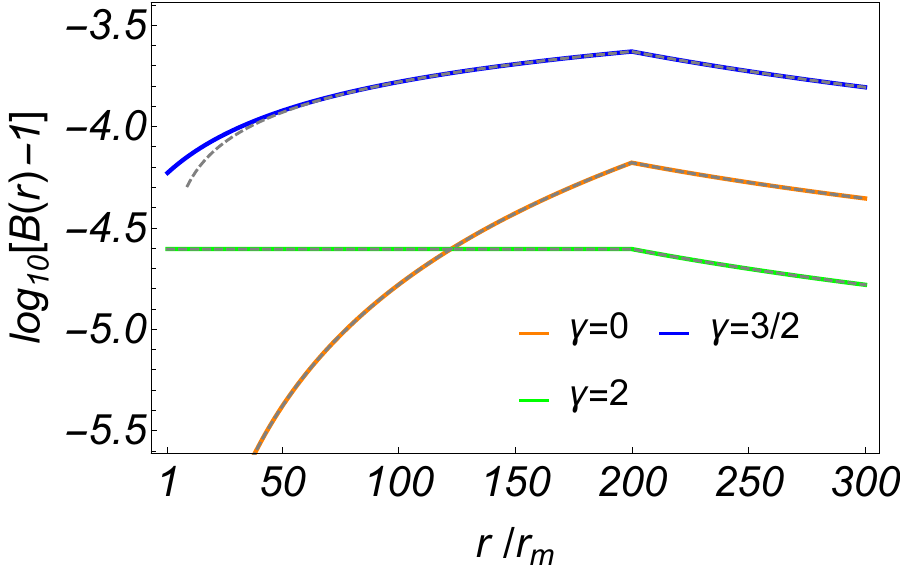}\\
(b)\\
\includegraphics[width=0.9\linewidth]{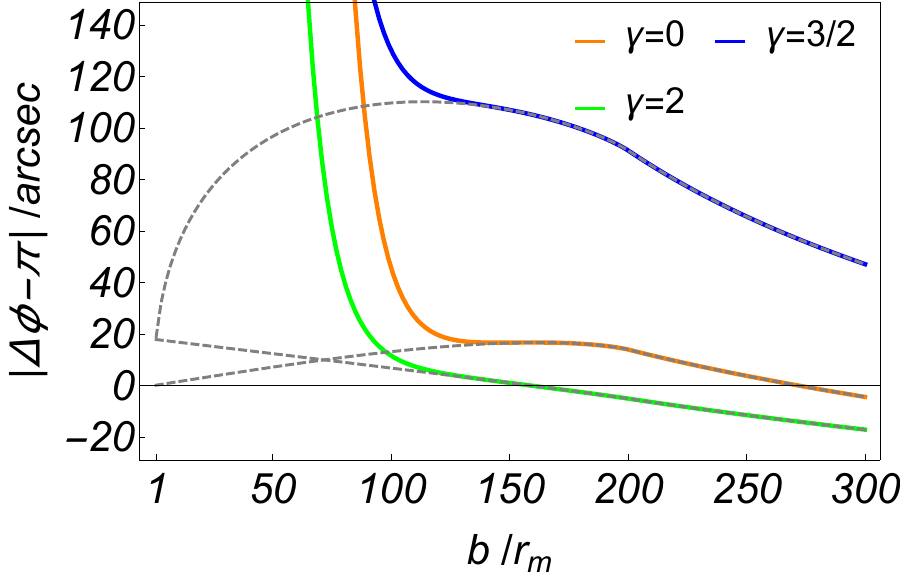}\\
(c)\\
\caption{\label{figmmmpowerlowplot}  
Perturbative (dashed lines) and numerical (solid lines) solutions of the metric functions and deflection of the power-law model with finite boundaries. $\gamma=0$ (uniform density), $\gamma=3/2$ and $\gamma=2$ (SIS density). The parameters used are $r_{s,d}=5\times10^{6} r_m$, $r_m=18~\mathrm{kpc}$ and $R=200 r_m$. For the density, $\rho_c=8\times10^{4}~M_{\odot}/\mathrm{kpc}^3$ for $\gamma=0$, and $\rho_c=4\times10^{8}~M_{\odot}/\mathrm{kpc}^3$ for $\gamma=3/2$ and $\gamma=2$. The smaller $\rho_c$ for $\gamma=0$ ensures that $M$ and consequently the deflection is comparable to the other two cases.
}
\end{figure}
Fig. \ref{figmmmpowerlowplot} shows the metric functions and the deflection angle for the uniform, SIS, and $\gamma=3/2$ densities, assuming the density boundary is at $R=200r_m$. Similar to the PIS model case, the non-smooth transitions at the boundary are also characteristic of the first-order matching condition. The metric functions and the deflection angle obtained using our series (solid lines) closely match the numerical results (dashed lines) for $b$ greater than approximately $R/2$. 

\section{Gravitational Lensing\label{secmmmgl}}

With the deflection angles for various density profiles established, the GL equation can be straightforwardly formulated, enabling the calculation of the apparent angles and magnifications of the lensed images.

When the deflection angle $\Delta\phi(r_0,~r_s,~r_d)$ includes the finite distance effects of the source and detector, as shown in our results in Eqs. \eqref{eqmmmdphiexpand}, \eqref{eqmmmgNFWexpand}, \eqref{eqmmmdphihm}, \eqref{eqmmmg3d}, \eqref{eqmmmg4d}, \eqref{eqmmmnonintegerdp}, \eqref{eqmmmdefpis} and \eqref{eqmmmdefpower}, the GL equation, which is essentially the definition of the deflection angle, takes the form given in
\begin{align}
    \Delta\phi(b,~r_s,~r_d)-\pi=\phi_s-\pi\equiv \delta\phi, \label{eqmmmgeneralleq} 
\end{align}
where $\phi_s$ is the $\phi$-coordinate of the source, and we set the detector's $\phi$-coordinate to zero without loss of generality. Here, $\delta\phi$ represents a small quantity that characterizes the deviation of the source from the lens-detector axis. The next step is to solve for $r_0$ from Eq. \eqref{eqmmmgeneralleq}, after which the apparent angles can be determined. 

\subsection{GL in density profiles extended to infinity}

\begin{figure}[htp!]
    \centering
    \includegraphics[width=0.9\linewidth]{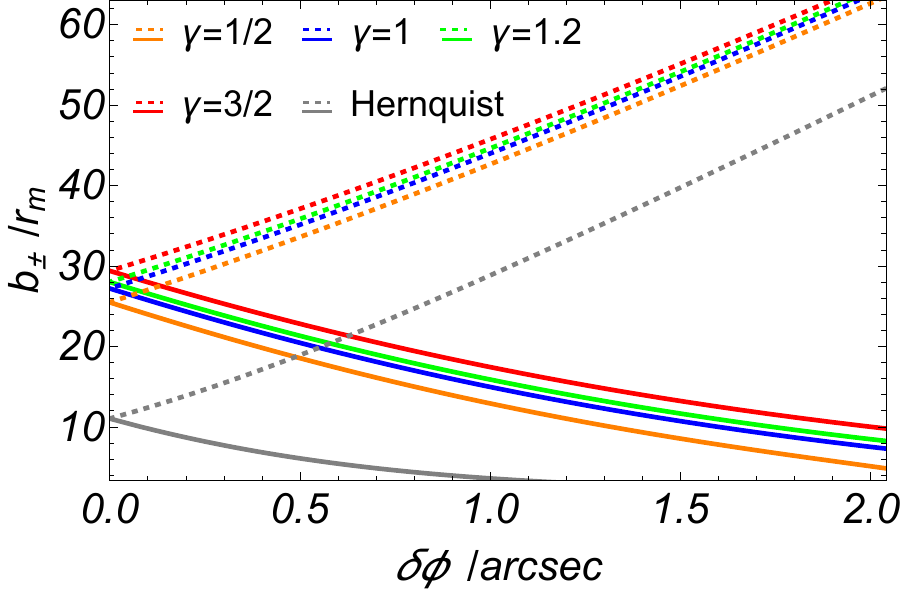}\\
    (a)\\
    \includegraphics[width=0.9\linewidth]{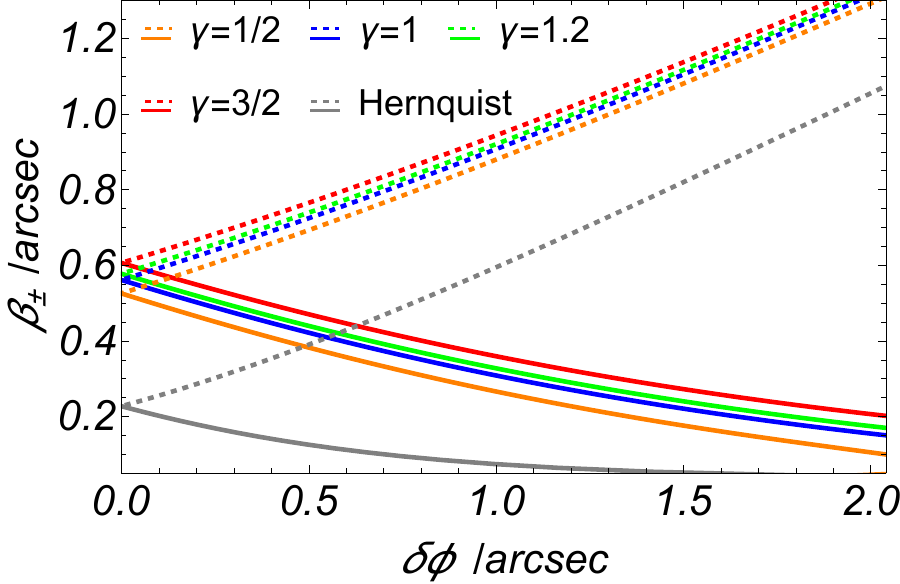}\\
    (b)\\
    \includegraphics[width=0.9\linewidth]{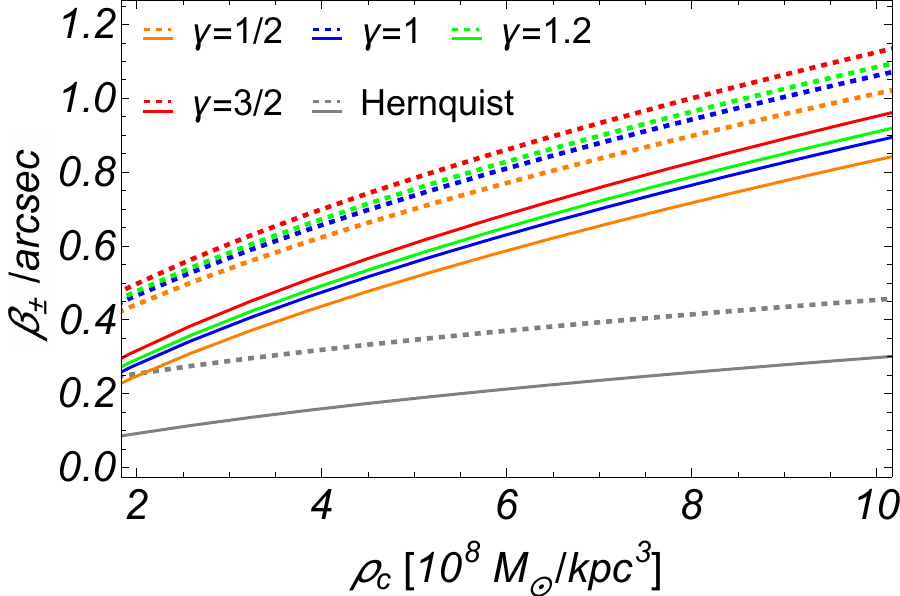}\\
    (c)
\caption{\label{figmmmgnfwhbbeta} Dependence of $b_\pm$ (plot (a)) and $\beta_\pm$ (plot (b)) on $\delta\phi $ in the gNFW and Hernquist models. Here we used Eq. \eqref{eqmmmaahern}. The parameter values for $\rho_c,\,r_{s,d},\, r_m,\,v$ are the same as in Figs. \ref{figmmmgnfwnfwplot}-\ref{figmmmhernquistplot}. (c) Variation of $\beta_\pm$ with $\rho_c$. Here $\delta\phi=0.3^{\prime\prime}$ and other parameters except $\rho_c$ are the same as those in (a) and (b). The solid and dashed curves represent counterclockwise ($+$) and clockwise ($-$) rotations, respectively. }
\end{figure}

Since $\Delta\phi(b,~r_s,~r_d)$
has been expanded as a series in $(b/r_{s,d})$, as shown in Eqs. \eqref{eqmmmdphiexpand}, \eqref{eqmmmgNFWexpand} and \eqref{eqmmmdphihm}, which are rational functions of $b$, it is straightforward to see that Eq. \eqref{eqmmmgeneralleq} can be transformed into a (quasi-)polynomial in $b$. The general GL equation in this case is
\begin{align}\label{eqmmmbanddphis}
    &\pi+\frac{1}{b}\left[b_{1,0}+b_{1,1}-\frac{a_{1,0}}{v^2}+\left(b_{1,1}-\frac{a_{1,1}}{v^2}\right)\ln{\frac{b}{2}}\right]\nonumber\\
    &-b\left(\frac{1}{r_s}+\frac{1}{r_d}\right)=\pi\pm \delta\phi. 
\end{align}
When $b_{1,1}=a_{1,1}=0$, this reduces to a quadratic equation commonly encountered in simple SSS spacetimes. In general, however, this equation can only be solved numerically, and we demonstrate that there are always two physical solutions at relatively large $b$. We denote the solutions as $b_\pm$, with $b_+$ (or $b_-$) representing the impact parameter of the counterclockwise (or clockwise) rotating trajectory (see Fig. \ref{figmmmschdiag} for illustration). In Fig. \ref{figmmmgnfwhbbeta} (a), we show how $b_\pm$ depends on the small source position angle $\delta\phi$ for the gNFW $(\gamma=0.5,\,1.5)$, NFW $(\gamma=1.0)$, and Hernquist models. It is observed that, for the gNFW and NFW models, $b_\pm$ increases slightly as $\gamma$ increases. This is consistent with the observation in Fig. \ref{figmmmgnfwnfwplot} that a larger $\gamma$ corresponds to a denser lens and a larger $m(r)$, resulting in stronger deflection, as shown in Fig. \ref{figmmmgnfwdef}. Moreover, the Hernquist model has a smaller $b_\pm$ compared to the gNFW or NFW models, which aligns with the observation in Fig. \ref{figmmmgnfwdef} and Fig. \ref{figmmmhernquistplot} (c), showing that it consistently has larger deflection angles for all considered $\gamma$ values. 

By substituting $b_\pm$ into Eq. \eqref{eqmmmbeta} and expanding for small $1/r_{s,d}$, we obtain the apparent angles of the two lens images as
\begin{align}\label{eqmmmapparent}
    \beta_{\pm}=&b_{\pm}\left[\frac{1}{r_{s,d}}+\frac{a_{1,0}+a_{1,1}\ln{r_{s,d}}}{2v^2 r_{s,d}^2}+\mathcal{O}\left(\frac{\ln^2{r_{s,d}}}{r_{s,d}^3}\right)\right]\nonumber\\
    =&b_{\pm}\left[\frac{1}{r_{s,d}}-\frac{m_0+4\pi\rho_3(1+\ln{r_{s,d}})}{v^2 r_{s,d}^2}+\mathcal{O}\left(\frac{\ln^2{r_{s,d}}}{r_{s,d}^3}\right)\right].
\end{align}
For the gNFW and Hernquist models, substituting the coefficients from Eqs. \eqref{eqmmmabninfint} and \eqref{eqmmmhmser}, respectively, these apparent angles are given by
\begin{align}
    \beta_{\pm,\mathrm{gNFW}}=&\frac{b_{\pm}}{r_m}\left[\frac{r_m}{r_{s,d}}+\frac{M}{r_m}\frac{2}{v^2}\left(C_{\gamma}-1-\ln{\frac{r_{s,d}}{r_m}}\right)\left(\frac{r_m}{r_{s,d}}\right)^2\right.\nonumber\\
    &\left.+\mathcal{O}\left(\frac{r_m^3}{r_{s,d}^3}\ln^2{\frac{r_{s,d}}{r_m}}\right)\right],~~\gamma>0,~\gamma\neq1, \label{eqmmmaagnfw}\\
    \beta_{\pm,\mathrm{H}}=&\frac{b_{\pm}}{r_m}\left[\frac{r_m}{r_{s,d}}-\frac{M}{r_m}\frac{1}{v^2}\left(\frac{r_m}{r_{s,d}}\right)^2+\mathcal{O}\left(\frac{r_m}{r_{s,d}}\right)^3\right]. \label{eqmmmaahern}
\end{align}
In Fig. \ref{figmmmgnfwhbbeta} (b), we plot the apparent angle $\beta_{\pm}$ from Eq. \eqref{eqmmmaagnfw} as a function of $\delta\phi$ for the gNFW model with $\gamma=0.5,\,1,\,1.2,\,1.5$, and from Eq. \eqref{eqmmmaahern} for the Hernquist model. We observe that as $\delta\phi$ increases from $0$ to $2^{\prime\prime}$, $\beta_{-}$ increases while $\beta_{+}$ decreases, which is consistent with general expectations and with $b_\pm$ shown in panel (a). 

To extract information about the density from the observables, we need to further investigate how the apparent angles $\beta_\pm$ depend on the galaxy halo size $r_m$ and the overall density scale $\rho_c$. For $r_m$, we observe from the density Eq. \eqref{eqmmmmmbf} that it always appears in conjunction with $r$ and vice versa. Therefore, it is straightforward to anticipate its effect on both $b_\pm$ and $\beta_\pm\approx b_\pm/r_d$. In the weak deflection limit (WDL), both quantities scale linearly with $r_m$. This observation is consistent with the behavior described in Eqs. \eqref{eqmmmaagnfw} and \eqref{eqmmmaahern}. For the overall density scale $\rho_c$, we further investigate its effect in Fig. \ref{figmmmgnfwhbbeta} (c). It is observed that the apparent angles $\beta_\pm$ of both images generally increase as $\rho_c$ increases, which is expected since a larger $\rho_c$ results in a greater mass inside a sphere with a fixed radius.  
Although this increase is not linear, it aligns with the general effect of a larger total mass $M$ on the apparent angles. Indeed, in the Schwarzschild case, the two apparent angles are given by \cite{Liu:2015zou}
\begin{align}
    \beta_{\pm,\mathrm{Sch}}= \frac{1}{2} \left( \sqrt{\delta \phi ^2+\frac{16 M r_i}{r_f
   \left(r_f+r_i\right)}}\mp \delta \phi \right).
\end{align}
It is straightforward to verify that these apparent angles exhibit a similar dependence on $M$ as on $\rho_c$ in Fig. \ref{figmmmgnfwhbbeta} (c).

\subsection{GL in densities with finite boundaries}

For the GL equation in densities with finite boundaries, we note that the deflection angles, as shown in Eq. \eqref{eqmmmdefpis} for the PIS model and Eq. \eqref{eqmmmdefpower} for the power-law models, inherently account for the finite distance effects of the source and detector. Therefore, similar to the case discussed in the previous subsection, these deflections naturally lead to a GL equation 
\begin{align}
\Delta\phi_{\mathrm{Sch}}+\frac{4 a_0 \sqrt{b_0} \left(a_0 \left(v^2-1\right)+1\right)}{2 a_0^2 \left(v^2-1\right)+2 a_0-a_1 R}\arccos{\frac{b}{b_R}}=\pi\pm\delta\phi.
\end{align}

From this, the desired impact parameters can be solved. We note that the coefficients $a_0, ~a_1$, and $b_0$ are provided in Eqs. \eqref{eqmmmabcoeffR} and are explicitly shown in the PIS and the power-law models in Eqs. \eqref{eqmmmdefpis} and \eqref{eqmmmdefpower}, respectively. 
However, since this equation involves polynomial, root, and $\arctan$ functions of $b$, we will solve it numerically.
Substituting the solved $b_{\pm}$ values into Eq. \eqref{eqmmmbeta}, the apparent angles of the two images are obtained as shown in 
\begin{align}
    \beta_{\pm,R}=&b_\pm\arcsin{\frac{1}{r_{s,d}}\sqrt{\frac{\lb 2M-r_{s,d}\rb v^2}{2M\lb v^2-1\rb-r_{s,d}v^2}}}. \label{eqmmmbetapmr}
\end{align}
It is worth noting that the additional factor alongside $b_\pm$ is the same as in the Schwarzschild case with mass $M$ and detector distance $r_d$ for both types of models. This is because the detector is located outside the density boundary, so the local spacetime at the detector effectively behaves like Schwarzschild geometry. 

\begin{figure}[htp!]
    \centering
    \includegraphics[width=0.9\linewidth]{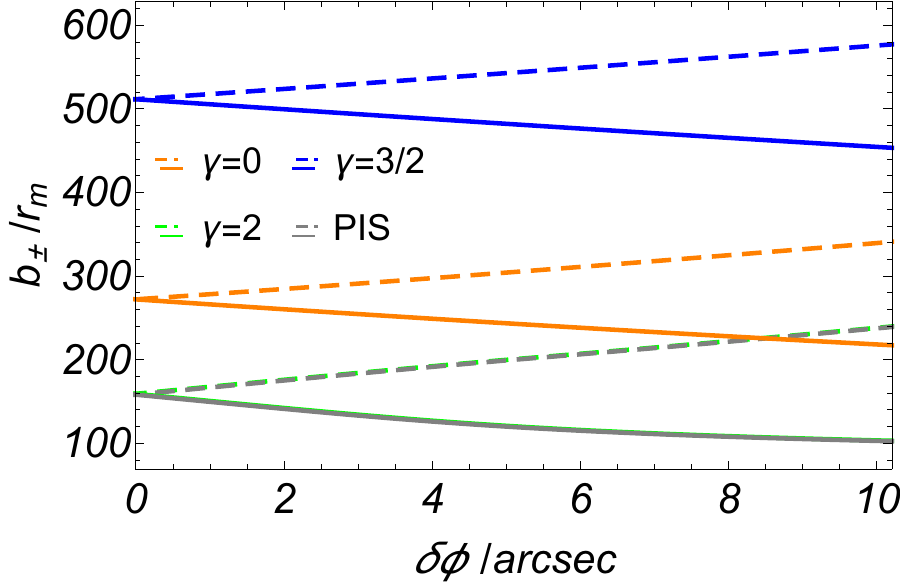}\\
    (a)\\
    \includegraphics[width=0.9\linewidth]{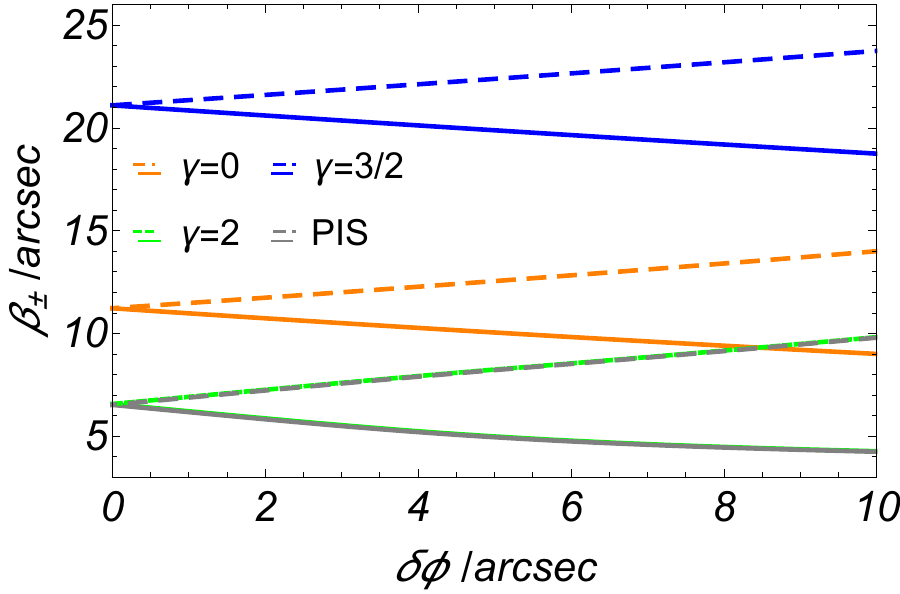}\\
    (b)\\
    \includegraphics[width=0.9\linewidth]{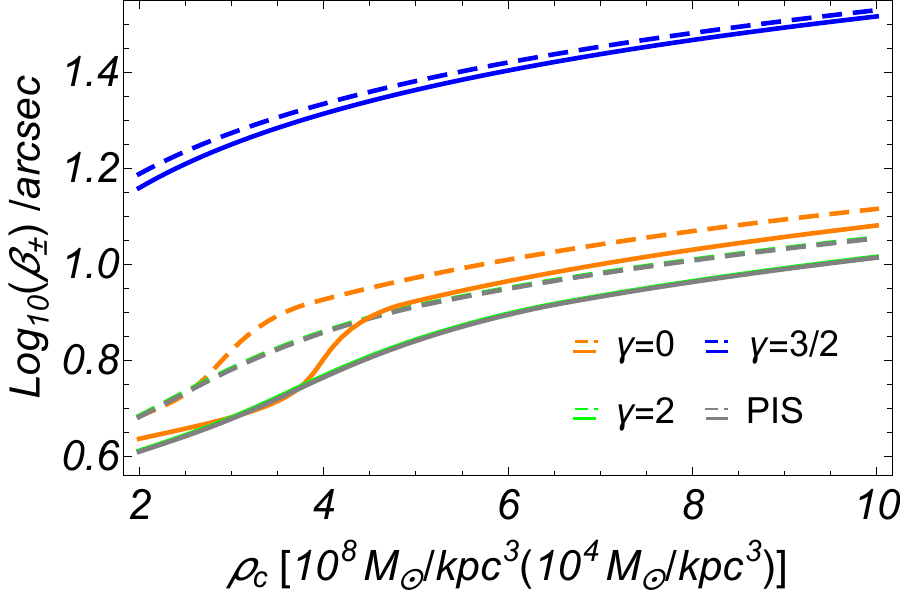}\\
    (c)\\
\caption{\label{figmmmfbap} Dependence of $b_\pm$ (a), $\beta_\pm$ (b) on $\delta\phi$ and $\rho_c$ (c) in the PIS and power-law models with finite boundaries.  Here we used Eq. \eqref{eqmmmbetapmr}. In (a) and (b), the parameter values for the PIS model are the same as in Fig. \ref{figmmmPISplot}, and for the power-law model, they are the same as in Fig. \ref{figmmmpowerlowplot}. In (c), $\delta\phi=2^{\prime\prime}$ is used and all other parameters except $\rho_c$ are the same as in (a) and (b). The solid and dashed curves represent particles on clockwise or counterclockwise trajectories, respectively. In (c), the unit $10^4~M_\odot/\mathrm{kpc}^3$ is for $\gamma=0$ and $10^8~M_\odot /\mathrm{kpc}^3$ is for other $\gamma$'s and PIS model. }
\end{figure}

In Fig. \ref{figmmmfbap}, we plot the solved impact parameters and apparent angles of the two images for the PIS, uniform, SIS, and $\gamma=3/2$ power-law models as functions of the source angle $\delta\phi$ and the density scale $\rho_c$. The parameter choices are consistent with those used in Figs. \ref{figmmmPISplot} and \ref{figmmmpowerlowplot} for the deflection angles in these models. In Fig. \ref{figmmmfbap} (a), the impact parameters $b_\pm$ are observed to be larger than $\sim 100 r_m$. Compared to Fig. \ref{figmmmPISplot} (b) and Fig. \ref{figmmmpowerlowplot} (b), these impact parameters fall within the range where our perturbative deflection angles are accurate, making these solutions reliable. Furthermore, by comparing the $\gamma=3/2$ and $\gamma=2$ cases for the power-law model, it is observed that $b_\pm$ decreases as $\gamma$ increases, indicating that a denser core (with the same total mass) is more effective in deflecting the signal. Both $b_\pm$ and $\beta_\pm$ for the PIS model are indistinguishable from those of the SIS case in these plots. This is because the results for the metric functions and deflection angles are series expansions around $R=200r_m$, which is much larger than $r_m$. The deviation between the SIS and PIS models only becomes noticeable closer to $r_m$, which is not observable in these plots. The apparent image angles shown in Fig. \ref{figmmmfbap} (b) for these cases follow the same trend as $b_\pm$ in (a), consistent with Eq. \eqref{eqmmmbetapmr}, where $r_d$ determines the other factor and remains constant across these cases. Finally, Fig. \ref{figmmmfbap} (c) illustrates the dependence of the apparent angles on the matter density scale $\rho_c$. As $\rho_c$ increases, the total mass enclosed within $R$ also increases linearly, resulting in a more pronounced bending of the trajectories towards the center. Consequently, to ensure that the signal still reaches the detector, larger $\beta_\pm$ are required, which aligns with Eq. \eqref{eqmmmbetapmr}. 

\section{Conclusion and Discussions\label{secmmmdisc}}

In this work, we investigate the correlation between the density profile of SSS perfect fluid spacetimes, the deflection of signals, as well as their GL. The general TOV equations are solved perturbatively, using either an asymptotic expansion for rapidly diminishing matter densities or a series expansion near a finite boundary. The matter density models considered include the gNFW model (which encompasses the simple NFW model), power-law, SIS, Hernquist, and PIS models. For each model, we derive the corresponding metric functions, or equivalently, the mass function and the gravitational potential, and analyze the deflection of signals in terms of the density profile expansion coefficients. Both null and timelike signal deflections are treated uniformly, with the finite distance effects of the source and detector naturally incorporated. These deflections lead to an exact GL equation in the weak-field regime, from which the impact parameters required for the signals to reach the detector are determined. The corresponding apparent angles of the GL images were also provided for each of the aforementioned matter models, and the effects of various density parameters on the deflection and apparent angles of the GL images were analyzed.  

The method developed in this work, along with the corresponding results, demonstrates that it is possible to directly derive the deflection and GL apparent angles for a wide range of mass densities, at least within the framework of the perfect fluid and SSS approximations. For the gNFW model and its sub-models, our results enable the metric functions and deflection angles to be expressed directly in terms of the asymptotic or boundary expansion coefficients of the density functions.
This approach also suggests that it can be extended to investigate the deflection and GL phenomena in more complex density distributions. Moreover, it encourages us to apply the same approach used in the perturbative solution method to other quantities, such as the time delay in GL or phenomena associated with bounded orbits. We are actively exploring these directions as a part of our future work. 

\acknowledgments
P. Liu is supported by the
Undergraduate Training Programs for Innovation and Entrepreneurship of Wuhan University. 

\appendix

\section{Integrability of the expansions \label{secmmmappint}}

In this appendix, we demonstrate how to evaluate the integrals given in Eqs. \eqref{eqmmmiintres}, \eqref{eqmmmnonintegarintegrate} and \eqref{eqmmmindef}, as discussed in the main text. We also provide the explicit formulas for each of these integrals. 

Firstly, for Eq. \eqref{eqmmmiintres}
\begin{align}\label{eqmmmiintresC}
    I_{n,m}(\beta_i,b)=&\int_{\sin{\beta_i}}^1\lb\frac{u}{b}\rb^n\lb\ln{\frac{u}{b}}\rb^m\frac{\mathrm{d}u}{\sqrt{1-u^2}}\nonumber\\
    =&\frac{1}{b^n}\int_{\beta_i}^{\pi/2}\sin^n{\theta}\lb\ln{\frac{\sin{\theta}}{b}}\rb^m\mathrm{d}\theta
,~~i=s,d,
\end{align}
we can verify that it satisfies the following recurrence relation 
\begin{align}\label{eqmmmintegralnandm}
I_{n,m}(\beta_i,b)=&\frac{\sin^{n-1}{\beta_i}\cos{\beta_i}}{nb^n}\lb\ln{\frac{\sin{\beta_i}}{b}}\rb^m\nonumber\\&-\frac{m}{n}I_{n,m-1}(\beta_i,b)+\frac{n-1}{nb^2}I_{n-2,m}(\beta_i,b)\nonumber\\&+\frac{m}{nb^2}I_{n-2,m-1}(\beta_i,b),\nonumber\\
    &~~~~~~~~~~~~ n=2,~3,~\cdots;~m=1,~2,~\cdots .
\end{align}
This recurrence relation allows us to compute all higher-order integrals $I_{n,m}$ starting from the lowest-order ones $I_{0,0},\,I_{0,1},\,I_{1,0},\,I_{1,1}$ and $I_{2,0}$. The values of these integrals can be determined easily using known formulas 
\begin{subequations}\label{eqmmmfirstfewinm}
\begin{align}
     &I_{0,0}(\beta_i,b)=\frac{\pi}{2}-\beta_i,\\
     &I_{0,1}(\beta_i,b)=\frac{1}{2} \left[-\ln (2 b) \left(\pi -2 \beta
   _i\right)+\Im\left(\text{Li}_2\left(\mathrm{e}^{-2 i \beta
   _i}\right)\right)\right],\\
     &I_{1,0}(\beta_i,b)=\frac{\cos{\beta_i}}{b},\\
     &I_{1,1}(\beta_i,b)=\frac{1}{b}\left\{\cos{\beta_i}\lsb\ln{\lb\frac{\sin{\beta_i}}{b}\rb}-1\rsb-\ln{\lb\tan{\frac{\beta_i}{2}}\rb}\right\},\\
     &I_{2,0}(\beta_i,b)=\frac{1}{4b^2}\lb\pi-2\beta_i+\sin{2\beta_i}\rb.
\end{align}
\end{subequations}
For the finite distance case, we can expand $I_{n,m}(\beta_i,b)$ for small $1/r_i$ by first using Eq. \eqref{eqmmmbeta} to expand $\beta_i$ in this limit. For the first few orders, we have
\begin{align}
    \beta_i=&\frac{b}{r_i}+\frac{b}{r_i^2}\lb\frac{a_{1,0}}{2v^2}+\frac{a_{1,1}}{2v^2}\ln{r_i}\rb+\frac{b^3}{6r_i^3}\nonumber\\
    &+\frac{b}{r_i^3}\left[\frac{a_{2,0}-a_{1,0}^2}{2v^2}+\frac{3a_{1,0}^2}{8v^4}+\lb\frac{3 a_{1,1}^2}{8 v^4}-\frac{a_{1,1}^2}{2v^2}\rb\ln^2{r_i}\right.\nonumber\\
    &\left.+\lb\frac{3 a_{1,0} a_{1,1}}{ 4v^4}+\frac{ a_{2,1}-2 a_{1,0} a_{1,1}}{ 2v^2}\rb\ln{r_i}\right]+\mathcal{O}\lb\frac{\ln^3{r_i}}{r_i^4}\rb.
\end{align}
For the case of series expansion for densities with non-integer powers, we need to integrate Eq. \eqref{eqmmmnonintegarintegrate}
\begin{align}\label{eqmmmnonintegarintegrateC}
    I_{n,m}(\beta_i)=&\int_{\sin{\beta_i}}^1\frac{u^{n+m\delta}}{\sqrt{1-u^2}}\mathrm{d}u\nonumber\\
    =&\int_{\beta_i}^{\frac{\pi}{2}}(\sin{\theta})^{n+m\delta}\mathrm{d}\theta,\quad i=s,d,
\end{align}
where $n$ and $m$ are integers and $\delta$ is a reduced fraction. This can be integrated to yield
\begin{align}
    &I_{n,m}(\beta_i)=\frac{\sqrt{\pi}\Gamma(\frac{1+n+m\delta}{2})}{2\Gamma(\frac{2+n+m\delta}{2})}-\frac{\sin{\beta_i}^{1+n+m\delta}}{1+n+m\delta}\nonumber\\&\times_2F_1(\frac{1}{2},\frac{1+n+m\delta}{2},\frac{3+n+m\delta}{2},\sin{\beta_i}^2),
    \label{eqmmmnmdintres}
\end{align}
where $_2F_1$ represents the 2F1 hypergeometric function.

We next consider the integral given in Eq. \eqref{eqmmmindef}, where $n$ is a non-negative integer. It becomes
\begin{align}\label{eqmmmintegral}
    I_{n}(\beta _i)&\equiv\frac{1}{b^n}\int_{\sin{\beta_R}}^1\frac{(u-b/b_R)^n}{\sqrt{1-u^2}}\mathrm{d}u\nonumber\\
    =&\frac{1}{b^n}\sum_{k=0}^n C_n^k \lb-\frac{b}{b_R}\rb^{n-k}\int_{\sin{\beta_R}}^1\frac{u^k}{\sqrt{1-u^2}}\mathrm{d}u\nonumber\\
    =&\frac{1}{b^n}\sum_{k=0}^n C_n^k \frac{(k-1)!!}{k!!}\lb-\frac{b}{b_R}\rb^{n-k}\nonumber\\
    &\times\begin{cases} \displaystyle 
    \frac{\pi}{2}-\beta_R+\cos{\beta_R}\sum_{j=1}^{[k/2]}\frac{(2j-2)!!}{(2j-1)!!}\sin^{2j-1}{\beta_R},\\
    \displaystyle 
    \cos{\beta_R}\lsb1+\sum_{j=1}^{[k/2]}\frac{{(2j-1)!!}}{(2j)!!}\sin^{2j}{\beta_R}\rsb,\\
    \end{cases}
\end{align}
where the first and second lines are for even and odd $k$, respectively. The first few orders are 
\begin{subequations}
\begin{align}
    I_0(\beta_R)=&\frac{\pi}{2}-\beta_R,\\
    I_1(\beta_R)=&\frac{1}{b}\left[\cos{\beta_R}-\lb\frac{\pi}{2}-\beta_R\rb\frac{b}{b_R}\right],\\
    I_2(\beta_R)=&\frac{1}{4b^2}\left[\pi-2\beta_R+\sin{2\beta_R}-8\cos{\beta_R}\frac{b}{b_R}\right.\nonumber\\
    &\left.+2(\pi-2\beta_R)\lb\frac{b}{b_R}\rb^2\right],\\
    I_3(\beta_R)=&\frac{1}{12b^3}\left[-6\lb\pi-2\beta_R\rb\lb\frac{b}{b_R}\rb^3+36\cos{\beta_R}\lb\frac{b}{b_R}\rb^2\right.\nonumber\\&\left.-9\lb\pi-2\beta_R+\sin{2\beta_R}\rb\frac{b}{b_R}+9\cos{\beta_R}-\cos{3\beta_R}\right].
\end{align}
\end{subequations}
Higher-order terms can also be easily obtained.

\vspace{2cm}

\bibliographystyle{apsrev}
\bibliography{references}
\end{document}